\documentclass{article}

\usepackage{PRIMEarxiv}

\usepackage[utf8]{inputenc} 
\usepackage[T1]{fontenc}    
\usepackage{hyperref}       
\usepackage{url}            
\usepackage{booktabs}       
\usepackage{amsfonts}       
\usepackage{nicefrac}       
\usepackage{microtype}      
\usepackage{lipsum}
\usepackage{fancyhdr}       
\usepackage{graphicx}       
\graphicspath{{media/}}     
\usepackage{subcaption}
\usepackage{amsmath}
\usepackage{multirow}
\usepackage[table,xcdraw]{xcolor}
\usepackage{caption}
\usepackage{cite}

\title{Domain-Aware Quantum Circuit for QML}

\author{
  Gurinder Singh\textsuperscript{1},
  Thaddeus Pellegrini\textsuperscript{2},
  and Kenneth M. Merz, Jr.\textsuperscript{1,*} \\
  \textsuperscript{1}Center for Computational Life Sciences, Lerner Research Institute, \\
  Cleveland Clinic, Cleveland, Ohio 44106, United States \\
  \textsuperscript{2}IBM Quantum, IBM T.J. Watson Research Center, \\
  Yorktown Heights, NY 10598, United States \\
  \texttt{singhg12@ccf.org}, \texttt{Thaddeus.Pellegrini@ibm.com} \\
  \textsuperscript{*}Corresponding author: \texttt{merzk@ccf.org}
}

\begin{document}
\maketitle

\begin{abstract}

Designing parameterized quantum circuits (PQCs) that are expressive, trainable, and robust to hardware noise is a central challenge for quantum machine learning (QML) on noisy intermediate-scale quantum (NISQ) devices. We present a Domain-Aware Quantum Circuit (DAQC) that leverages image priors to guide locality-preserving encoding and entanglement via non-overlapping DCT-style zigzag windows. The design employs interleaved encode–entangle–train cycles, where entanglement is applied among qubits hosting neighboring pixels, aligned to device connectivity. This staged, locality-preserving information flow expands the effective receptive field without deep global mixing, enabling efficient use of limited depth and qubits. The design concentrates representational capacity on short-range correlations, reduces long-range two-qubit operations, and encourages stable optimization, thereby mitigating depth-induced and globally entangled barren-plateau effects. We evaluate DAQC on MNIST, FashionMNIST, and PneumoniaMNIST datasets. On quantum hardware, DAQC achieves performance competitive with strong classical baselines (e.g., ResNet-18/50, DenseNet-121, EfficientNet-B0) and substantially outperforming Quantum Circuit Search (QCS) baselines.  To the best of our knowledge, DAQC, which uses a quantum feature extractor with only a linear classical readout (no deep classical backbone), currently achieves the best reported performance on real quantum hardware for QML-based image classification tasks. Code and pretrained models are available at: https://github.com/gurinder-hub/DAQC.

\end{abstract}


\section{Introduction}

QML is an exciting and rapidly growing area at the intersection of quantum algorithms and modern machine learning \cite{lloyd2013quantum, Biamonte2017, schuld2018supervised, schuld2021machine, Cerezo_2022, review_2025}. QML aims to tackle complex real-world problems that challenge classical methods, seeking potential quantum advantage by leveraging uniquely quantum properties such as superposition and entanglement. In the NISQ era \cite{preskill2018quantum}, however, practical QML faces key obstacles: efficient encoding of high-dimensional features into quantum states, noise, decoherence, readout error, circuit depth, hardware connectivity constraints (lead to SWAP/compilation overhead), trainability issues including barren plateaus, and limited hardware access. Despite these challenges, the field has advanced rapidly in recent years, spanning variational and kernel-based methods \cite{mitarai2018quantum,farhi2018classificationquantumneuralnetworks,Havl_ek_2019,Schuld_2020,lloyd2020quantumembeddingsmachinelearning,Liu_2021,Abbas_2021,schuld2021effect,schuld2021quantum,kiani2020quantummedicalimagingalgorithms,xiao2023practical,Cong_2019,kollias2023quantum,fan2023hybrid,cherrat2024quantum,De_Lorenzis_2025} and scalable training/simulation frameworks like kernel-alignment optimization, hierarchical learning, and large-scale simulation \cite{Sahin_2024,chen2025validatinglargescalequantummachine,Gharibyan2024,gharibyan2025quantumimageloadinghierarchical,gharibyan2025quantumimageclassificationexperiments}.

Most QML approaches employ parameterized quantum circuits (PQCs) with data-agnostic entanglement patterns and generic inductive biases. Although expressive, these circuits do not explicitly encode domain knowledge; for imaging tasks, they typically ignore canonical regularities such as strong correlations between neighboring pixels, approximate shift consistency, and multiscale spatial statistics that underpin the success of classical CNNs and vision transformers. Under realistic depth and connectivity budgets, many PQCs can embed only a limited number of input features (or rely on costly data re-uploading strategy), and non-native couplings induce SWAPs that further increase error. Combined with finite-shot readout and barren-plateau effects, these factors often degrade predictive performance. It is also worth noting that many of the QML works rely on hybrid quantum-classical models. In such settings, the individual contribution of the quantum and classical parts is often unclear. Moreover, claims of quantum advantage typically appear in restricted regimes (small datasets) with limited evidence of cross-domain generalization or reproducibility under realistic hardware constraints. 

Lack of well-supported guidelines for QML circuit design further motivates the community to adopt QCS frameworks such as QuantumSupernet \cite{du2022quantum}, QuantumNAS \cite{hanruiwang2022quantumnas}, Élivágar \cite{anagolum2024elivagarefficientquantumcircuit}, and QuProFS \cite{gujju2025quprofs} to automatically find noise-robust circuits for QML, but they come with practical drawbacks. First, their multi-stage pipelines including supernet training, evolutionary co-search of circuit topology and qubit mapping, evaluation under noise models, and post-hoc pruning are computationally expensive and time-consuming, often requiring many circuit evaluations and large shot budgets before a single deployable circuit emerges. Secondly, even with noise awareness, these approaches still optimize over large generic search spaces that frequently begin data-agnostic. Therefore, domain priors such as image locality and multiscale structure remain underexploited during circuit construction. Also, these frameworks score candidate circuits using proxy metrics (for example, clifford noise-resilience or circuit expressivity). These scores do not always predict how well a circuit will run on real hardware. Lastly, the initial search space is usually device-agnostic in these approaches and the method later has to co-optimize the qubit layout for a specific quantum hardware. This mapping results in additional SWAP gates to satisfy connectivity, which increases compiled depth and error rates.

To address these limitations, we propose a DAQC that integrates the image-domain priors-especially correlations among neighboring pixels with the practical constraints of NISQ hardware. We use non-overlapping, DCT-style zigzag windows so that spatial neighbors are encoded sequentially on adjacent qubits. The circuit runs interleaved encode–entangle–train cycles: periodic, local ECR gates couple neighboring qubits aligned with hardware connectivity, reducing long-range interactions and two-qubit error. This interleaved hierarchy incrementally spreads and consolidates information, enabling multi-stage feature aggregation under a controlled two-qubit budget. Our design targets useful expressivity aligned with the image manifold, improving optimization stability and mitigating barren-plateau risks. Experiments on standard image-classification datasets show strong performance and generalization relative to competitive classical baselines and QCS frameworks. By using a pure quantum circuit with only a linear classical readout, we enable clearer attribution of quantum contributions and establish a strong quantum baseline for image classification. 

The paper is organized as follows. Section~2 introduces the proposed DAQC, analyzes its expressibility and entangling capability, and describes the circuit implementation, training, and inference details. Section~3 presents the experimental results and the barren plateau analysis. Section~4 provides concluding remarks and outlook toward quantum advantage.

\section{Domain-aware quantum circuit}

\begin{figure}[!ht]
    \centering
    \includegraphics[width=\linewidth]{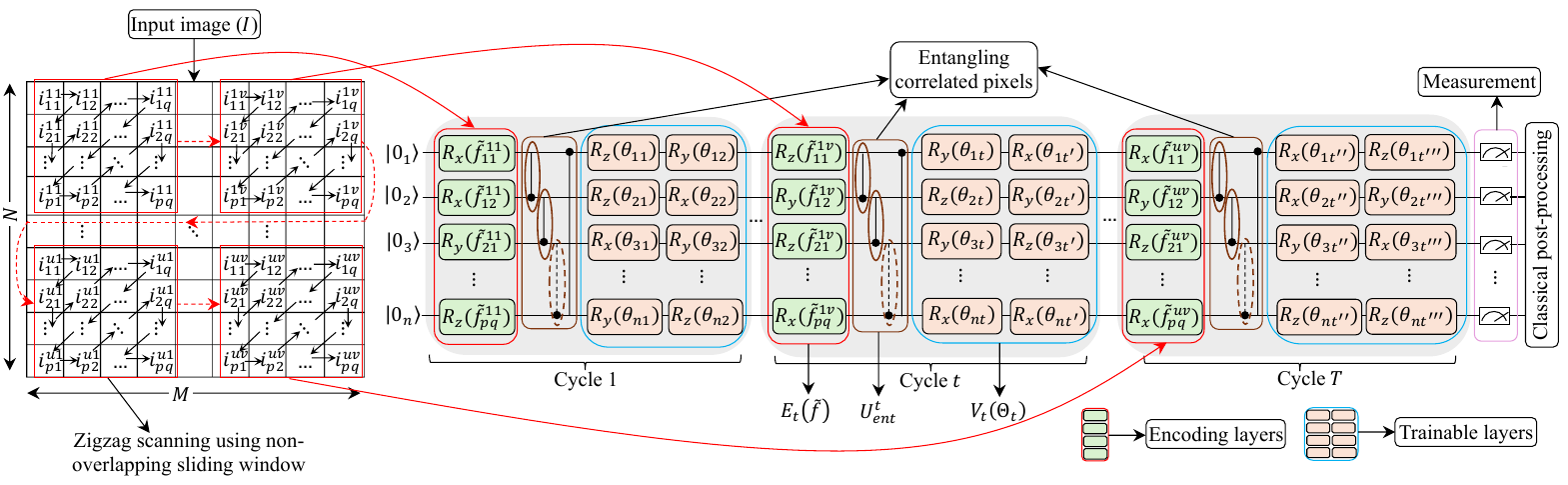}
    \caption{Domain-aware quantum circuit for QML. Here $p\times q$ is the size of non-overlapping sliding window for the zigzag scan, $u\times v$ denotes the total number of non-overlapping sliding windows. In the trainable layers, $t=2v-1$, $t'=2v$, $t''=2(u\times v)-1$, and $t'''=2(u\times v)$. Encoding and trainable layers gates are selected based on uniform random sampling from $RXYZ$ design space.}
    \label{figure1}
\end{figure}

The domain-aware quantum circuit encodes image structure by mapping neighboring pixels to adjacent qubits and coupling them locally as illustrated in Fig. \ref{figure1}. Due to current hardware constraints, we first downsample the input image \(I\) to a compact grid \(P(I)\in\mathbb{R}^{N\times M}\) via adaptive average pooling to match the qubit budget and reduce noise exposure. It is worth noting that downsampling results in information loss and can potentially deteriorate the representational capacity of the quantum circuit as compared to the classical baselines capable of processing the full-resolution images. The pooled grid is then partitioned into non-overlapping \(p\times q\) patches. Each patch is traversed in a DCT-style zigzag order, producing a (\(p\times q\))-vector \({z}_{u,v}=[i_{11}^{uv},i_{12}^{uv},i_{21}^{uv},...,i_{pq}^{uv}]\) such that \(u=N/p\) and \(v=M/q\). All the resultant vectors after zigzag traversal are concatenated in raster order resulting in an $NM$-dimensional feature vector \(
{f}(I)
=\big[\,{z}_{1,1}\,\|\,{z}_{1,2}\,\|\,{z}_{1,3}\,\|\,\cdots\,\|\,{z}_{u,v}\,\big]
\in\mathbb{R}^{NM}
\). We used angle encoding in our quantum circuit design to map the classical data to quantum states. To achieve this, we map the raw intensities to angles in \([0,\pi]\) using affine normalization
\(
\tilde{{f}}_k \;=\; \pi \cdot ({{f}_k - \min({f})})/({\max({f}) - \min({f})}), k=1,2,\ldots,NM
\). The normalization stabilizes gradients and ensures the encoding rotation angles lie in a natural range. This normalized feature vector facilitates the sequential encoding of neighboring correlated pixels onto adjacent qubits via Pauli rotation gates as shown in Fig. \ref{figure1}.

The circuit is built as a stack of interleaved cycles; each cycle performs, in order, (i) feature encoding, (ii) local entanglement, and (iii) trainable one-qubit rotations. Interleaving cycles prevent long runs of data-only or parameter-only layers, improving gradient flow and mitigating barren-plateaus. Entanglement is applied between qubits hosting neighboring pixels using hardware-friendly two-qubit gates (ECR) aligned with device connectivity, which minimizes long-range interactions and two-qubit error exposure while explicitly modeling pixel adjacency. Repeating this cadence across patches spreads and consolidates information without deep global mixing. The unitary realized by the circuit can be written compactly as: 

\begin{equation}
U(\tilde{{f}},\Theta) \;=\; \prod_{t=1}^{T} V_t(\Theta_t)\, U_{ent}^{(t)}\,E_t(\tilde{{f}})\,
\end{equation}

where \(\Theta\) represents the trainable rotation parameters, \(T \;=\; \left\lceil \frac{NM}{\, n} \right\rceil\) is the number of interleaved cycles in the quantum circuit. In cycle $t$, the embedding operator is a tensor product of single-qubit rotations defined as: 

\begin{equation}
\label{eq:formal-U}
E_t(\tilde{{f}})
= \bigotimes_{q=1}^{n} R_{\sigma_{t,q}}\!\big(\tilde f_{k(t,q)}\big),
\qquad k(t,q) = (t-1)n + q
\end{equation}

where $n$ is the number of qubits, \(\sigma_{t,q}\) is the axis used by the embedding rotation on qubit $q$ during cycle $t$, and $k(t,q)$ denotes the feature index in the normalized vector $\tilde{{f}}$. The QuantumNAS circuit design framework \cite{hanruiwang2022quantumnas} explored 6 different design spaces including $U3+CU3$, $ZZ+RY$, $RXYZ$, $ZX+XX$, $RXYZ+U1+CU3$, and $IBMQ Basis$ to design quantum circuits. The experiments conducted in QuantumNAS \cite{hanruiwang2022quantumnas} shows that overall the $RXYZ$ design space outperforms the other design spaces. Motivated by this, we explored the $RXYZ$ design space in our quantum circuit design. We sampled $\sigma_{t,q}$ from the $RXYZ$ domain space (i.e., $R_{\sigma_{t,q}}$ is $R_x$, $R_y$, or $R_z$) using uniform random sampling, encouraging symmetry breaking and stabilizing optimization. This small injection of stochasticity has a large practical effect: (i) it promotes gradient isotropy so updates are not stuck along one axis, (ii) it reduces accidental commutation/alignment that can zero out local derivatives, and (iii) it acts as implicit regularization, often improving generalization without extra depth.

The circular nearest-neighbor ring entanglement pattern is consistent with pixel adjacency, maps well onto heavy-hex–like device topologies, and minimizes SWAP insertion by the transpiler. In each interleaved cycle, whenever we apply an entangling block, we therefore use a nearest-neighbor ring to couple adjacent qubits. However, inserting an entangling layer in every cycle makes the circuit substantially deeper and two-qubit–gate heavy, which on NISQ hardware leads to increased noise accumulation and empirically degraded performance. To quantify this trade-off, we analyze the expressibility and entangling capability of DAQC in Section~2.1 and perform a complementary ablation study in Section~3.4 (Table~\ref{table3}), which indicates that the entangling layer should be inserted sparsely with a tunable period $f_{etn}$. The density of entanglement layers in the quantum circuit is therefore an explicit \emph{knob} that trades expressivity against noise: smaller $f_{etn}$ (more frequent entangling) increases mixing but also error exposure, whereas larger $f_{etn}$ (sparser entangling) reduces two-qubit error at the cost of weaker global coupling. This periodic, local coupling strategy injects multi-qubit correlations while controlling two-qubit error accumulation and compilation overhead. For cycle $t$ in our quantum circuit, the $U_{ent}^{(t)}$ is defined as:

\begin{equation}
U_{ent}^{(t)} =
\begin{cases}
\displaystyle \prod_{(i,j)\in\mathcal{E}} {ECR}_{i,j}, & \text{if } ((t-1)\bmod f_{etn})=0,\\[6pt]
\mathbb{I}, & \text{otherwise.}
\end{cases}
\end{equation}

where, \(\mathcal{E} = \{(1,2),(2,3),...,(n,1)\}\) denotes the set of neighbor pairs (ring edges). Each cycle includes exactly two trainable layers, each a tensor of axis-sampled 1-parameter rotations with trainable angles $\theta^{(k)}_{t,q}$ defined as:

\begin{equation}
\label{eq:Vt-Klayers}
V_t(\Theta_t)
= \prod_{k=1}^{2}
\left(\bigotimes_{q=1}^{n} R_{\tau^{(k)}_{t,q}}\!\big(\theta^{(k)}_{t,q}\big)\right)
\end{equation}

where \(\tau_{t,q}\in\{x,y,z\}\) is the axis used by the trainable rotation on qubit $q$ during cycle $t$ in trainable layer $k$ with $k\in \{1,2\}$ and is also sampled using uniform random sampling. With two trainable columns per cycle and $T$ cycles, the total number of variational angles is $2nT$. This interleaved hierarchy of encoding, local ECR, and trainable layers mirrors the classical vision models building features from local to coarse scales under strict depth constraints. The ECR layer creates short-range correlations while information is still localized; two lightweight trainable layers refine these correlations. Repeating this pattern grows the effective receptive field gradually: early cycles capture very local edges and corners; later cycles allow information to spread across neighboring patches without deep, global mixing. Mapping neighbor pairs to adjacent qubits keeps the design hardware-conscious, so compiled depth scales predictably with entanglement frequency and the number of cycles. After the final interleaving cycle of the quantum circuit, we measure all qubits in the Pauli-\(Z\) basis to obtain the output vector \(\mathbf{m}(I;\Theta)= \big(\langle Z_0\rangle,\ldots,\langle Z_{n-1} \rangle\big)\) with \(\langle Z_i\rangle=\big\langle 0\big|U^\dagger(\Theta)\, Z_i \,U(\Theta)\big|0\big\rangle\). We map these expectations to logits in the classical post-processing step via a linear readout defined by \(
\ell(I) \;=\; W\,\mathbf{m}(I;\Theta) + \mathbf{b}\), which feed a cross-entropy loss defined as:

\begin{equation}
\mathcal{L}(\Theta)
= -\frac{1}{B}\sum_{i=1}^{B}
\log\!\left(
\frac{\exp\!\big(\ell_{y_i}^{(i)}\big)}
{\sum_{c=1}^{C}\exp\!\big(\ell_{c}^{(i)}\big)}
\right)
\end{equation}

where, $B$ represents the number of samples in the current mini-batch, $C$ is the number of classes, $\ell_{c}^{(i)}$ denotes the logit for class $c$ on sample $i$, and $y_i$ is the ground-truth class index for sample $i$. Although we measure all $n$ qubits, DAQC is interrogated only through the single-qubit Pauli-$Z$ expectation values $\{\langle Z_i\rangle\}_{i=1}^{n}$ and their classical linear combinations. At the operator level this corresponds to sums of 1-local observables $\sum_i Z_i$, rather than a single $n$-body operator such as $Z^{\otimes n}$. Consequently, the quantum contribution to the loss has the structure of a local cost function rather than a global cost. Local costs are known to exhibit significantly milder gradient scaling in $n$ compared to global costs that can lead to exponential barren plateaus \cite{Cerezo_2021}. To investigate this further, we performed barren plateau analysis of DAQC in Section 3.7.

The net effect of these design choices is a circuit that is simultaneously data-aware, hardware-aware, and training-aware. Data awareness is achieved by encoding and coupling precisely where images exhibit structure: zigzag ordering delivers adjacent pixels to adjacent wires, and the immediate local ECR layer captures their correlations at low cost. Hardware awareness comes from use of entanglement between adjacent qubits aligned with device connectivity, which minimizes transpiler-inserted SWAPs and lowers two-qubit error exposure. Training awareness follows from maintaining shallow cycles with explicit control over entanglement density and from using light stochasticity in rotation placement to keep gradients healthy. Together these properties produce what we call \emph{useful expressivity}: the circuit is expressive along directions that matter for the image manifold, without paying the depth and noise penalties of unnecessary global mixing. In practice, this alignment translates into better accuracy-per-two-qubit-gate on simulators and, crucially, into more faithful transfer of those gains to real hardware.

\subsection{Expressibility and entangling capability of the DAQC ansatz}

\begin{figure}[!t]
  \centering
  \begin{subfigure}[t]{0.33\linewidth}
    \centering
    \includegraphics[width=\linewidth]{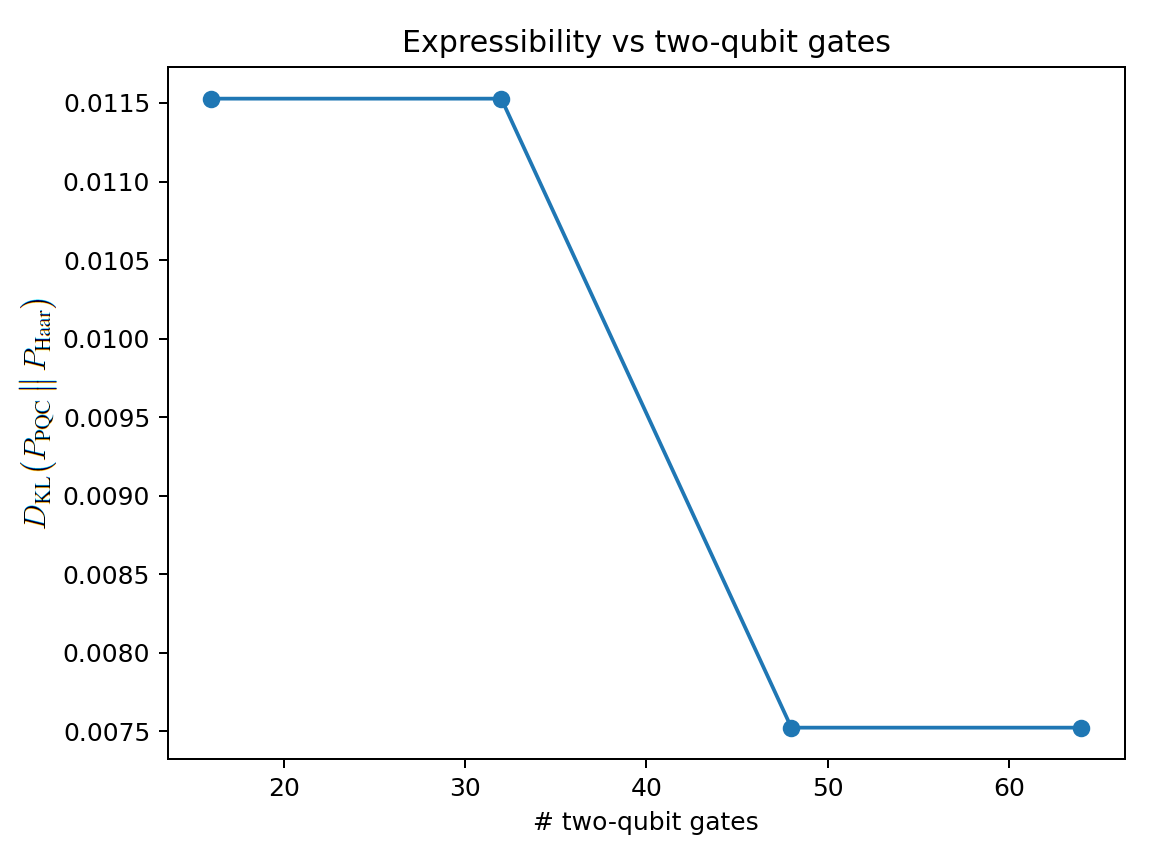}
    \caption{}
    \label{fig:daqc-a}
  \end{subfigure}\hfill
  \begin{subfigure}[t]{0.33\linewidth}
    \centering
    \includegraphics[width=\linewidth]{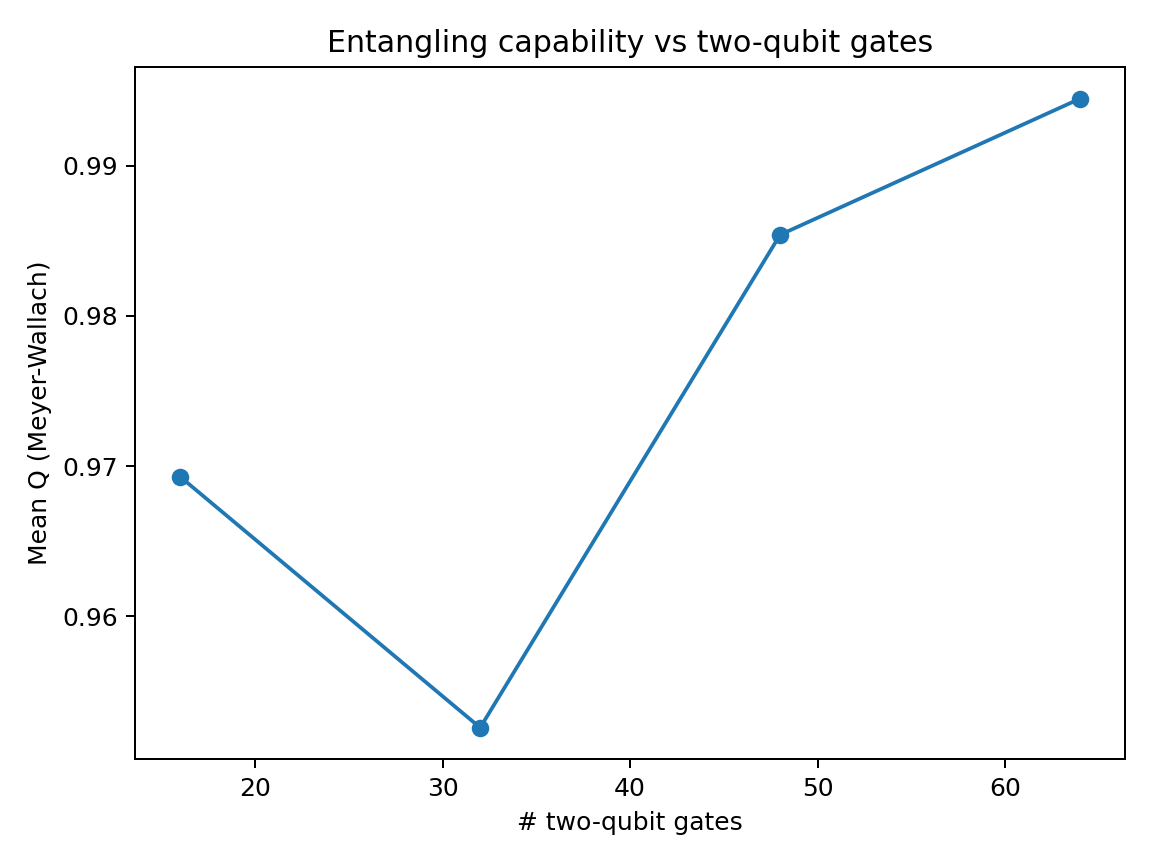}
    \caption{}
    \label{fig:daqc-b}
  \end{subfigure}\hfill
  \begin{subfigure}[t]{0.33\linewidth}
    \centering
    \includegraphics[width=\linewidth]{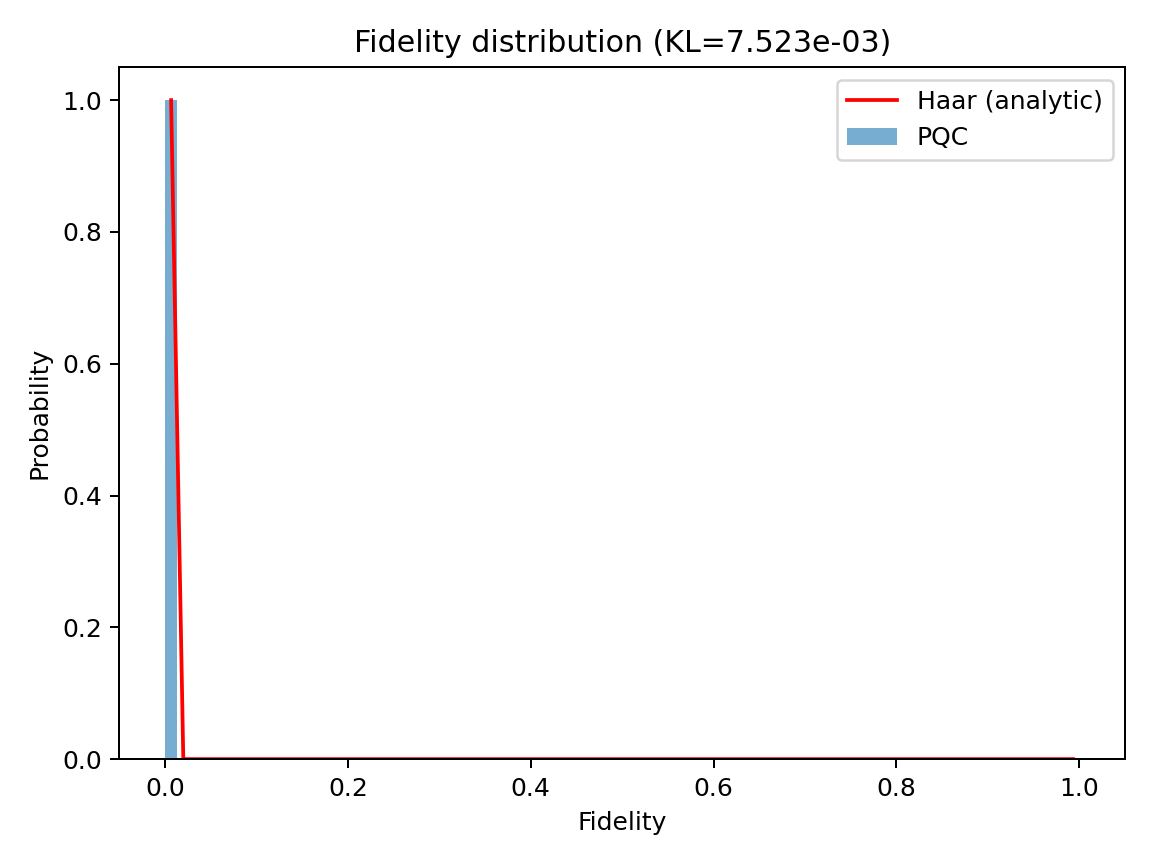}
    \caption{}
    \label{fig:daqc-c}
  \end{subfigure}
\caption{Expressibility and entangling capability of the 16-qubit DAQC. (a) Expressibility measured via $D_{KL}(P_{PQC}\Vert P_{Haar})$ (lower is better), (b) Mean Meyer--Wallach global entanglement measure $Q$ (higher is better), and (c) Fidelity histogram for the depth setting \((N_{embed}, N_{train}, N_{ECR}) \in (256,512,64)\) overlaid with the analytic Haar fidelity density.}
\label{figure2}
\end{figure}

To better understand the representational power of the DAQC ansatz, we follow the framework of Sim \emph{et al.}~\cite{Sim_2019} for quantifying the expressibility and entangling capability of parameterized quantum circuits. It is worth mentioning that we limit our DAQC design to 16 qubits in this study due to the limitations of current quantum hardware. In this analysis, we vary the number of embedding gates and trainable parameters in DAQC. We consider four depth settings with \((N_{embed}, N_{train}, N_{ECR}) \in \{(64,128, 16), (128,256,32), (192,384,48), (256,512,64)\}\) as shown in Figs. \ref{figure2}a and \ref{figure2}b, note that in the plots we mention only two-qubit gates along the x-axis for simplicity. These four settings correspond to total gate counts \(\{208, 416, 624, 832\}\). 

The expressibility of a parametrized quantum circuit is quantified by comparing the distribution of pairwise state fidelities generated by the ansatz to that of Haar-random states. For each depth setting, we first sample \(N_{states} = 2000\) random parameter vectors (both embedding angles and variational parameters drawn uniformly from \([0,\pi]\) and \([0,2\pi]\), respectively), simulate the corresponding statevectors on 16 qubits, and then estimate the fidelity distribution from \(N_{pairs} = 5000\) randomly chosen state pairs. The empirical histogram is compared against the analytic Haar
fidelity density defined by \(P_{Haar}(F) = \bigl(2^{n} - 1\bigr)\,(1 - F)^{2^{n} - 2}\) \cite{zyczkowski2005average} where $F$ represents fidelity, using the Kullback–Leibler divergence \(D_{KL}\!\left(P_{PQC} \,\big\Vert\, P_{Haar}\right)\) \cite{kullback1951information} which serves as our expressibility metric (lower value means high expressibility). For the shallowest circuit with 16 two-qubit gates, we obtain \(D_{KL} = 1.15 \times 10^{-2}\) as shown in Fig. \ref{figure2}a. Increasing the number of ECR gates to 32 leaves the KL divergence essentially unchanged, indicating that simply doubling the entangling depth at this stage does not substantially improve coverage of Hilbert space.
A more pronounced improvement is observed when going to 48 two-qubit gates, where \(D_{KL}\) drops to \(7.5 \times 10^{-3}\) showing high expressibility. Further increasing the depth to 64 two-qubit gates yields virtually identical expressibility, suggesting that the circuit has entered a \emph{saturation} regime in the sense of Sim \emph{et al.}~\cite{Sim_2019}: additional layers no longer significantly reduce the deviation from Haar, but do increase depth and parameter count.

We quantify entangling capability using the Meyer–Wallach global entanglement measure (\(Q\)) \cite{meyer2001global}, defined as the average linear entropy of all the single qubit reduced states. For a given depth setting, we reuse the same ensemble of randomly
sampled states and compute the mean of \(Q\) over that ensemble. As discussed in Sim \emph{et al.}~\cite{Sim_2019}, for Haar-random states on \(n\) qubits, the mean \(Q\) = \((2^{n} - 2)/(2^{n} + 1)\), which is very close to 1 for \(n = 16\). The dependence of the mean \(Q\) on the different depth settings is reported in Fig. \ref{figure2}b. Even the shallowest configuration already produces strongly entangled states, with mean \(Q \approx 0.97\). Interestingly, increasing the number of entanglers from 16 to 32 slightly reduces the mean entanglement to \(Q\approx 0.95\). This behavior is consistent with the general picture in
Sim~\emph{et al.}~\cite{Sim_2019}, where different circuit
families exhibit distinct trends in mean $Q$ as a function of
depth and the relationship between expressibility and entangling
capability is not one-to-one; circuits with similar mean $Q$ can display noticeably different expressibility and vice versa. As the depth is further increased to 64 two-qubit gates, the mean $Q$ rises to \(Q = 0.9944\), indicating that the circuit predominantly explores highly entangled regions of the state space and approaches the Haar-random benchmark.

Collectively, the expressibility and entangling-capability curves show that the DAQC reaches a regime of Haar-like behavior in both descriptors for the 64 two-qubit gates depth setting on 16 qubits. Beyond this point, additional layers will change the KL divergence and mean \(Q\) marginally, while circuit depth, parameter count, and hardware noise overhead continue to grow. Therefore, \((N_{\text{embed}}, N_{\text{train}}, N_{\text{ECR}}) \in (256,512,64)\) is an optimal case scenario for a 16 qubit circuit as it achieves near-saturated expressibility and entangling capability while keeping the two-qubit
gate count within the budget of current NISQ devices. Moreover, the fidelity histogram for the more expressive configuration (64 two-qubit gates) is shown in Fig. \ref{figure2}c showing that the empirical distribution closely tracks the analytic Haar curve. This confirms that the DAQC ansatz, at moderate depth, is able to generate a set of states whose pairwise overlaps are statistically very similar to those of Haar-random states on 16 qubits, i.e., it is highly expressible without requiring an excessively deep circuit. From a design perspective, this analysis provide a concrete guideline for choosing the depth of the DAQC circuit in our classification experiments. 

\subsection{Circuit implementation, training, and inference details}

In our final circuit configuration, we set the number of qubits = 16, zigzag scanning window size $4\times 4$ resulting in 256 embedding features for each input image, number of trainable parameters = 512, and $f_{etn}$ = 4 (supported by the expressibility and entangling capability analysis in Section 2.1 and ablation study in Section 3.4), which yields \(T \;=\; \left\lceil \frac{256}{\, 16} \right\rceil=16\) interleaved cycles; with 0-based indexing and the condition $((t-1)\bmod f_{etn})=0$, the ECR layers occur at human-readable cycles $t$={1,5,9,13}. The pixel-to-qubit mapping, choice of rotation gates, and column ordering follow the DAQC design. We use the Adam optimizer with a learning rate of 0.005, a weight decay of 0.0001, a cosine-annealed scheduling over 250 epochs with a batch size of 64, and early stopping (patience = 20) on AUC validation. Due to hardware constraints, we performed the training of quantum circuit models on a classical NVIDIA A100 GPU (43 GB RAM) using the TorchQuantum noiseless simulator \cite{hanruiwang2022quantumnas} and inference is done on the \texttt{ibm\_kingston} hardware. This approach is suitable in the NISQ era where full on-hardware training over larger datasets is impractical. For all the quantum experiments, the original input images of size $28\times 28$ are downsampled to $16\times 16$ using adaptive average pooling.

For real hardware evaluation, we keep the learned parameters fixed and run inference on \texttt{ibm\_kingston}. For each dataset considered, we randomly selected 200 samples from the testing set for evaluation on the real device due to the limited access. Moreover, based on the ablation study (Section 3.4) on different error suppression and mitigation techniques of SamplerV2 and EstimatorV2, we used Qiskit EstimatorV2 in Session execution mode by enabling an error-mitigation stack comprising of Dynamical Decoupling (DD) \cite{Niu2024MultiQubitDD}, Pauli Twirling (Twir) \cite{Kim2023ScalableErrorMitigation}, Twirled Readout Extinction (TREX) \cite{Nation2021ScalableMeasurementMitigation}, and Zero Noise Extrapolation (ZNE) \cite{Majumdar2023BestPracticesZNE} to reduce the effect of hardware noise. We note that Error Mitigation techniques (eg. ZNE) are only compatible with the Estimator Qiskit Primitive, used in this study, but not with the Sampler. We used 32000 shots for all the hardware experiments. 

\subsubsection{Transpilation details}

 A crucial step in running quantum workflows on real quantum hardware is optimizing the logical circuit for the target device during circuit transpilation. Transpilation \cite{transpiler_qiskit} has two key phases: layout and routing. Layout selects and places the physical qubits that will host the logical qubits. Routing handles the connectivity constraints between those physical qubits by introducing SWAP operations between the physical qubits that are not directly connected. Because a SWAP typically decomposes into three CNOTs, transpilers often explore multiple candidate layouts and prefer those that minimize SWAPs.

\begin{figure}[!h]
    \centering
    \includegraphics[width=0.7\linewidth]{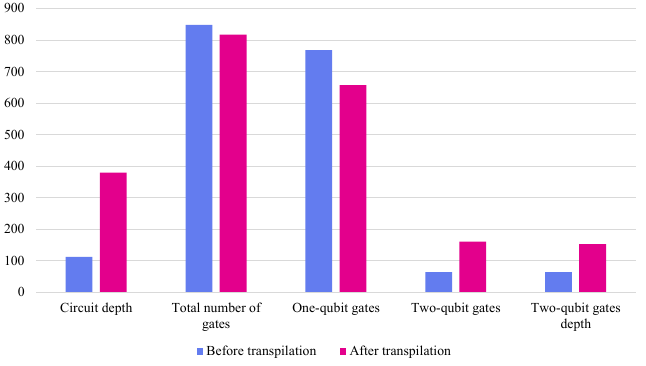}
    \caption{Specifications of DAQC before and after transpilation.}
    \label{figure3}
\end{figure}

In our study, a quantum circuit was created for each testing sample and transpiled using the optimization level 3 to control the circuit depth and gate count. Optimization level 3 is the most abstracted use of the Qiskit Transpiler, allowing a user to avoid implementation details around layout selection by handling them in an automated fashion. Moreover, optimization of 3 includes layout and routing optimization via an under-the-hood heuristic SWAP optimizer. Fig. \ref{figure3} shows the DAQC specifications before and after transpilation using the Qiskit transpile service. The Qiskit transpiler preserves overall gate budget but increasing the entangling cost and depth due to routing and scheduling on the target heavy-hex topology. The total gate count drops slightly from 848 to 818 driven by 1-qubit gate cancellations/merges i.e., from 768 to 657, but the 2-qubit gate count rises from 64 to 161 and 2-qubit depth from 64 to 153. Thus, the total circuit depth grows from 113 to 380. This pattern is typical for hardware mapping: the transpiler removes redundant single-qubit frames while inserting additional entangling operations (e.g., due to SWAP routing) to satisfy connectivity. 

\begin{figure}[!h]
  \centering
  \begin{subfigure}[t]{0.23\linewidth}
    \centering
    \includegraphics[width=\linewidth]{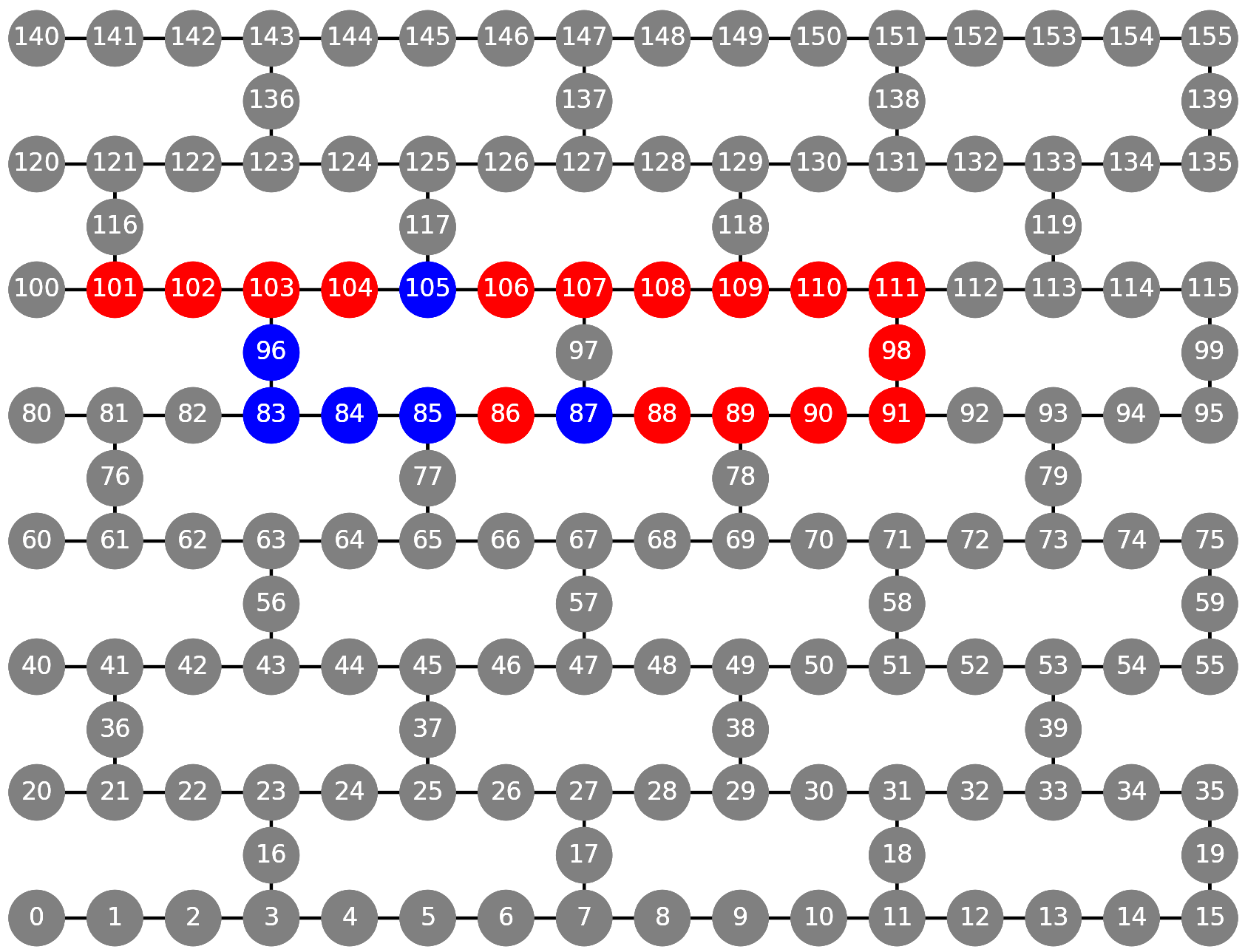}
    \caption{}
    \label{fig:daqc-a}
  \end{subfigure}\hfill
  \begin{subfigure}[t]{0.23\linewidth}
    \centering
    \includegraphics[width=\linewidth]{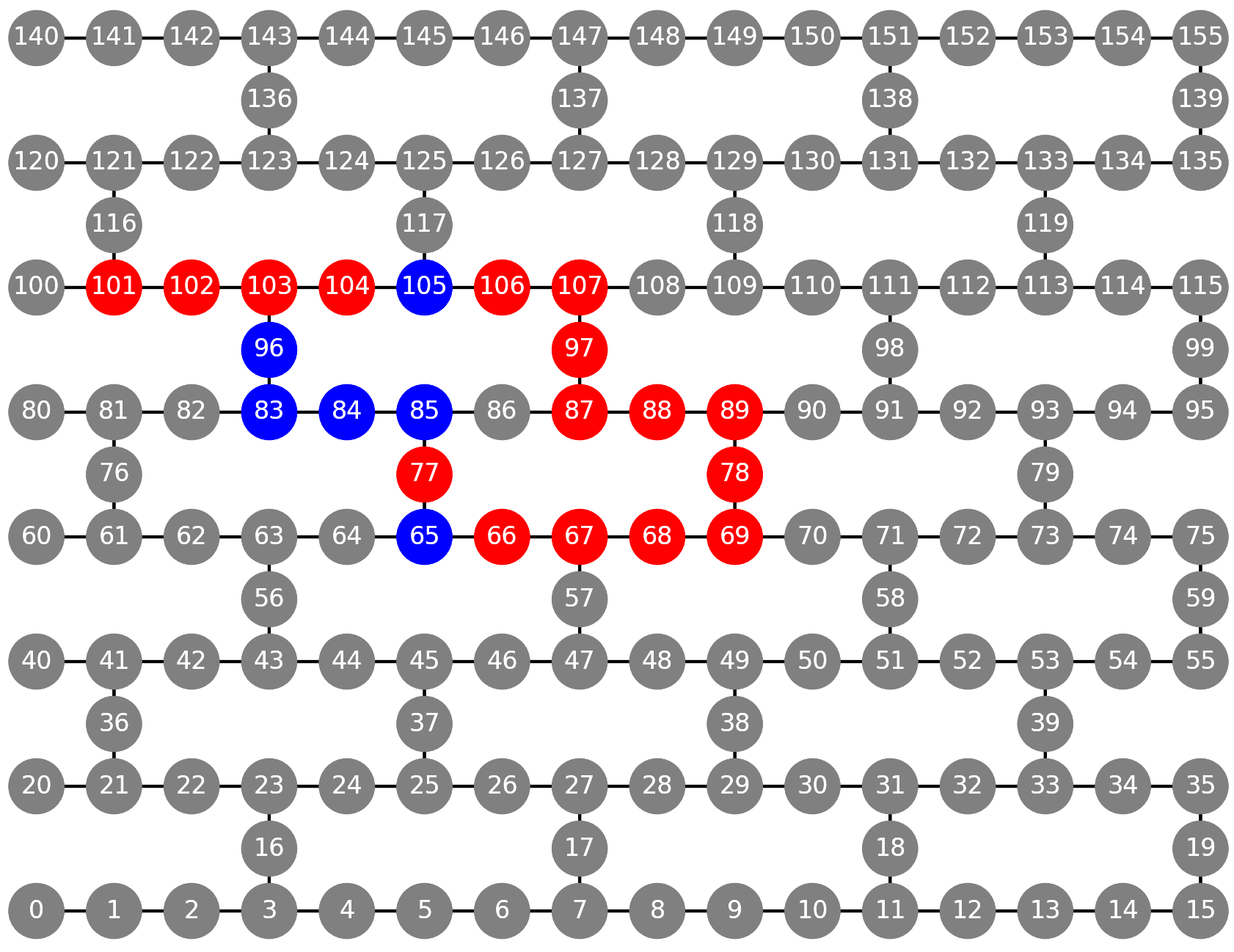}
    \caption{}
    \label{fig:daqc-b}
  \end{subfigure}\hfill
  \begin{subfigure}[t]{0.23\linewidth}
    \centering
    \includegraphics[width=\linewidth]{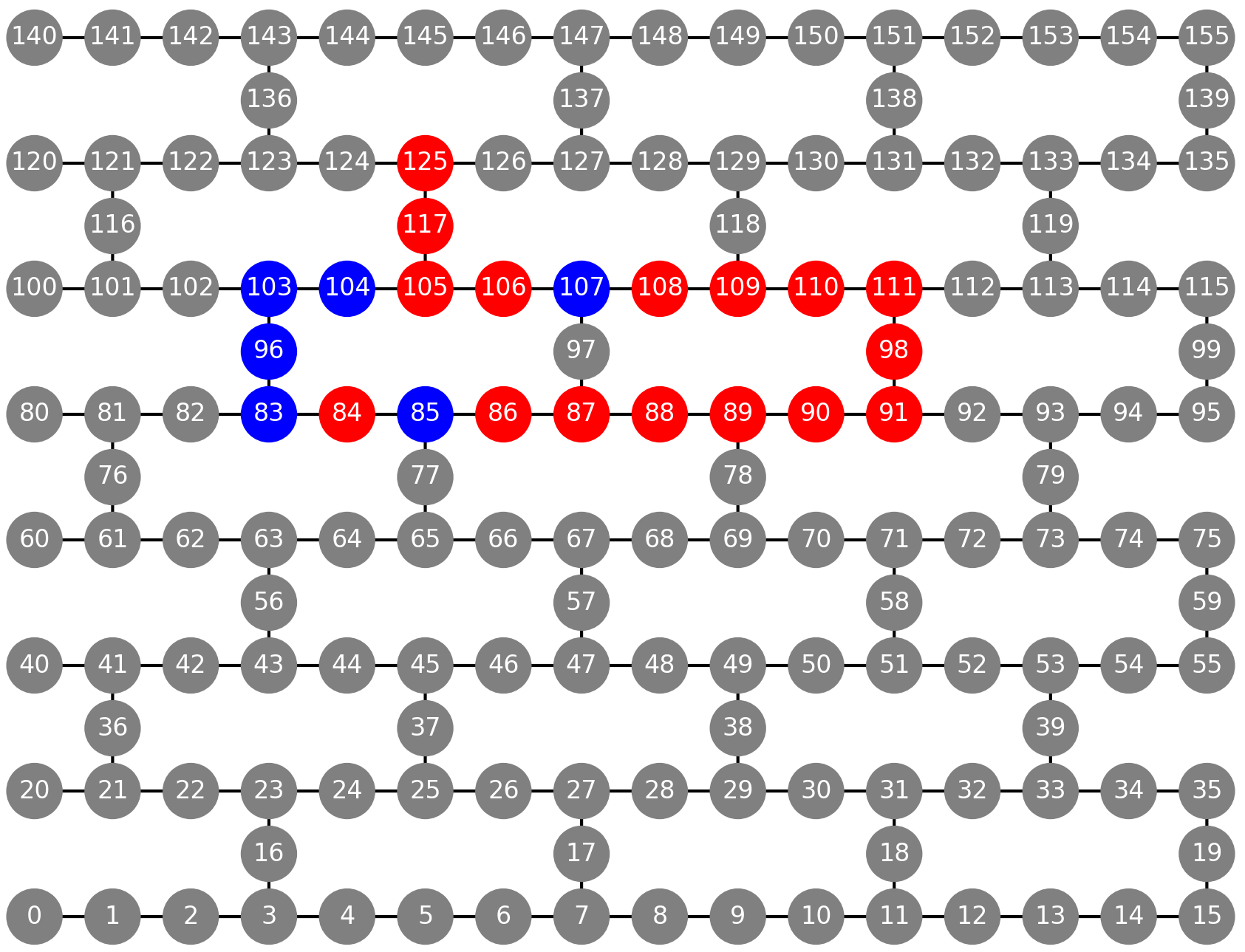}
    \caption{}
    \label{fig:daqc-c}
  \end{subfigure}\hfill
 \begin{subfigure}[t]{0.23\linewidth}
    \centering
    \includegraphics[width=\linewidth]{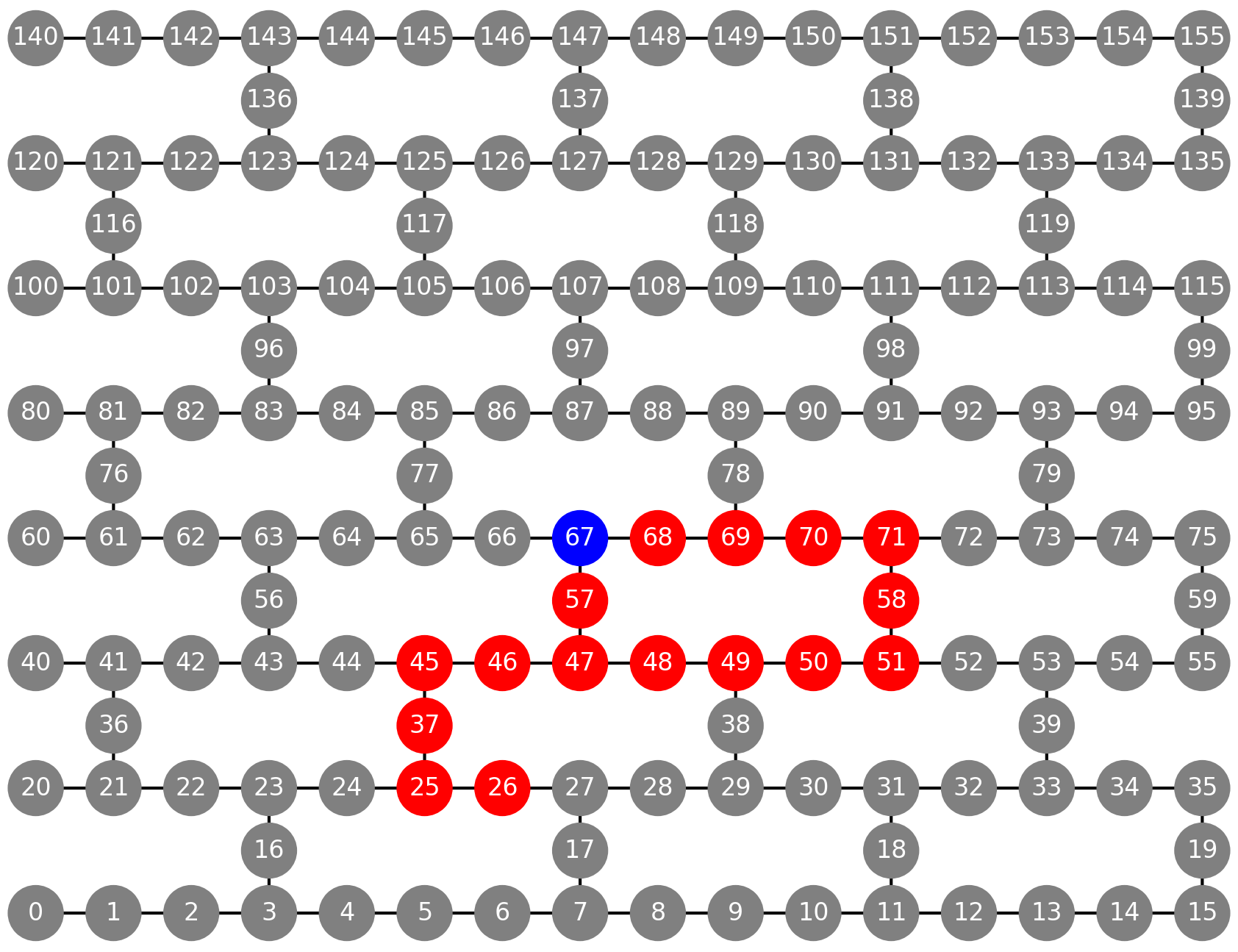}
    \caption{}
    \label{fig:daqc-c}
  \end{subfigure}
\caption{Some of the example layouts selected by the Qiskit transpiler for the DAQC on the \texttt{ibm\_kingston} hardware, red: physical qubits, blue: ancilla qubits.}
\label{figure4}
\end{figure}

Fig. \ref{figure4} shows some example layouts selected by the Qiskit tanspiler service for our DAQC on the \texttt{ibm\_kingston} hardware. In the resulting physical circuit, additional qubits may be introduced as ancilla qubits to bridge sparse connectivity—especially on superconducting processors with a heavy-hex lattice like those provided by IBM Quantum and used in this study (\texttt{ibm\_kingston}). These ancilla, shown in Fig. \ref{figure4} in blue, primarily relay quantum information (e.g., serve as waypoints for SWAP paths) and are not measured at the end of the protocol. For instance, the transpiled circuit occupied 22 physical qubits, of which 6 were ancilla needed to facilitate intermediary quantum operations between disconnected physical qubits via SWAP operations as shown in Figs \ref{figure4}a, \ref{figure4}b, and \ref{figure4}c. 

\subsubsection{Error profile of IBM Kingston quantum hardware}

\begin{figure}[!h]
  \centering
  \begin{subfigure}[t]{0.5\linewidth}
    \centering
    \includegraphics[width=\linewidth]{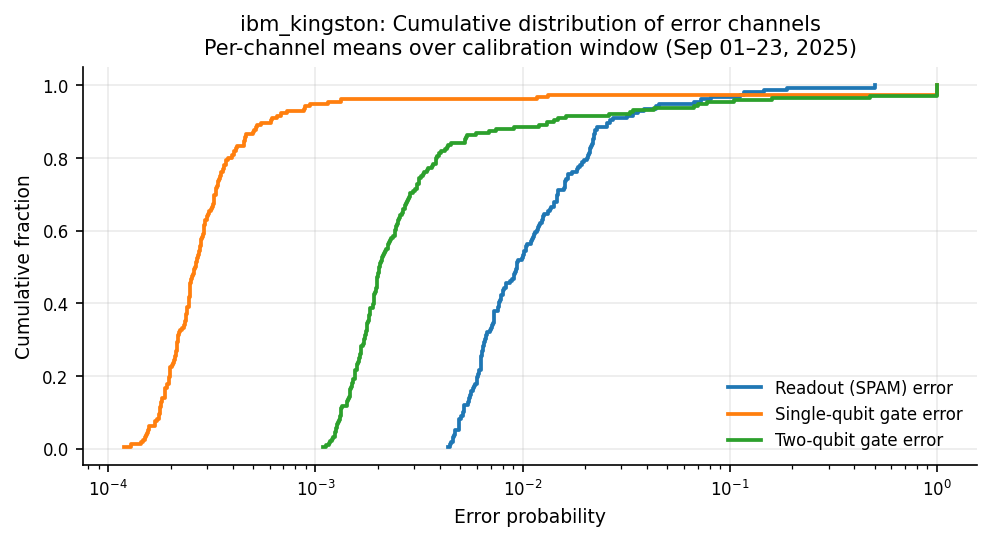}
    \caption{}
    \label{fig:daqc-a}
  \end{subfigure}\hfill
  \begin{subfigure}[t]{0.5\linewidth}
    \centering
    \includegraphics[width=\linewidth]{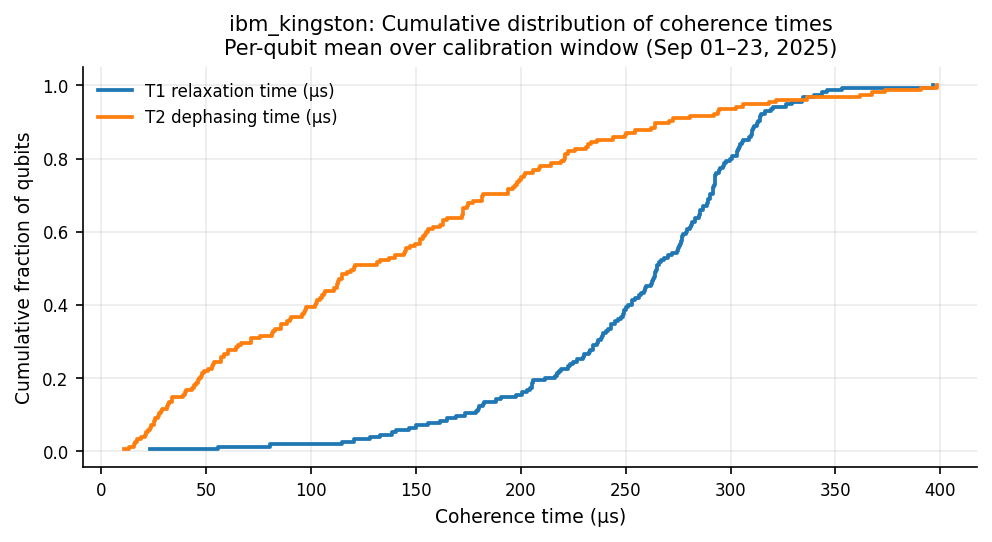}
    \caption{}
    \label{fig:daqc-b}
  \end{subfigure}\hfill
\caption{Hardware characterization data reported from the IBM Kingston quantum processor averaged across time on each error channel during the time-frame of September 01-23, 2025.}
\label{figure5}
\end{figure}

We provide the quantum hardware error profile of the \texttt{ibm\_kingston} QPU during the experimentation window (Sep 01-23, 2025) as shown in Fig. \ref{figure5}. During the experimentation window, 7 session jobs were executed with one job per dataset. We considered 200 samples or circuits for each dataset. In total, we ran about 1400 circuits on the \texttt{ibm\_kingston} quantum hardware during this experimentation window using the Estimator Qiskit Primitive incorporating a combination of error mitigation techniques (DD+TREX+Twir+ZNE). Each circuit execution with 32000 shots takes around 41 seconds of QPU time. 

Fig. \ref{figure5}a shows that single-qubit gate error was remarkably low during the experimentation window, with a median error of 3e-4. The two-qubit gate error was an order of magnitude more severe than one-qubit error, with a median of 2e-3 and readout error dominated with a median of 1e-2. The devices were characterized via Randomized Benchmarking \cite{Magesan2011ScalableRB}. The median T1 relaxation time was 260 $\mu$s, while the median T2 dephasing time was 130 $\mu$s as shown in Fig. \ref{figure5}b. \cite{Krantz2019QuantumEngineersGuide}

\section{Experimental results}

\subsection{Datasets}

We evaluated DAQC on three different standard image datasets having images of size 28$\times$28 including MNIST \cite{deng2012mnist}, FashionMNIST \cite{xiao2017fashion} , and PneumoniaMNIST \cite{medmnistv2} to validate the performance. The MNIST dataset contains 70,000 grayscale images (60,000 for training and 10,000 for testing) of handwritten digits from 0 to 9. Inspired by Élivágar \cite{anagolum2024elivagarefficientquantumcircuit}, we created 3 subsets from the full MNIST dataset including MNIST-2 (using classes {0, 1}), MNIST-4 (using classes {0, 1, 4, 8}), and complete MNIST-10 for evaluation. The FashionMNIST dataset is a direct drop-in replacement for the original MNIST containing 70,000 grayscale images (60,000 for training and 10,000 for testing) related to the fashion products of 10 different categories. Similarly to MNIST, we created 3 subsets from the complete FashionMNIST dataset including FashionMNIST-2 (using classes {0, 1}), FashionMNIST-4 (using classes {0, 1, 8, 9}), and complete FashionMNIST-10. To further evaluate the generalizability of our quantum model, we considered another challenging 2-class PneumoniaMNIST dataset containing 4708, 524, and 624 grayscale Chest X-Ray images for training, validation, and testing, respectively. This PneumoniaMNIST dataset is quite imbalanced with a training set class distribution of 1214 and 3494 samples. All available images are used in the evaluation for all the considered datasets. For all subsets of the MNIST and FashionMNIST datasets, 20\% images are used for validation from the available training subset and testing is performed on all available images in the test set. It is important to mention that while the noiseless results are evaluated on all the available test images, the quantum hardware evaluation only considered randomly selected 200 testing samples for all the considered datasets. However, we noticed that in all the considered datasets, performance metrics obtained using 200 samples are comparable to the whole testing set. The performance of the considered classical and QML models is measured using well-known classification metrics including Area Under the Curve (AUC), Accuracy (ACC), Specificity, Sensitivity, and F1-score.

\subsection{Classical baselines}

The objective of this work is to explore a potential quantum advantage over the classical baselines in terms of performance and model complexity. To validate this, we selected well-known standard classical ML baselines used for image classification including ResNet18 \cite{he2016deep}, ResNet50 \cite{he2016deep}, DenseNet121 \cite{huang2018denselyconnectedconvolutionalnetworks}, and EfficientB0 \cite{tan2020efficientnetrethinkingmodelscaling}. In all classical ML baselines, we used a standard model backbone without pretraining adapted to 28$\times$28 inputs using CIFAR-style configuration (the first convolution layer is replaced with a convolution layer of kernel size 3$\times$3, stride 1, and padding 1) for small-sized images as mentioned in the baseline ResNet paper \cite{he2016deep}. Also, the initial max-pooling layer is removed to prevent spatial down-sampling and last a fully-connected linear layer is modified based on the number of classes in the dataset. The preprocessing includes the conversion of grayscale images to RGB images (using channel duplication), normalized based on mean and standard deviation. For the MNIST data set, we used a mean of 0.1307 and a standard deviation of 0.3081 applied per channel after duplication. For the FashionMNIST and PneumoniaMNIST datasets, we used a mean and standard deviation of 0.5 to normalize the input images. All of these classical baselines are trained for 100 epochs with a batch size of 128, Adam optimizer (learning rate of 0.001, weight decay of 0.0001), and a multi-step learning scheduler (milestones at 50\% and 75\% of training and $\gamma$ of 0.1). The cross-entropy loss function is used on logits to calculate the loss. 

\subsection{Competing QCS methods}

We also compared DAQC with the state-of-the-art (SOTA) QCS methods including QuantumSupernet \cite{du2022quantum}, QuantumNAS \cite{hanruiwang2022quantumnas}, Élivágar \cite{anagolum2024elivagarefficientquantumcircuit}, and QuProFS \cite{gujju2025quprofs}. We selected the MNIST-10 dataset to provide a more robust comparison between these methods because using a simple variant like MNIST-2 will not effectively evaluate the model capabilities. All of these SOTA QCS methods are open-source except QuProFS. We ran the QuantumSupernet, QuantumNAS, and Élivágar frameworks using preprocessing and hyperparameter settings consistent with their original baseline papers. As we evaluated our DAQC on \texttt{ibm\_kingston} hardware, we used the \texttt{ibm\_kingston} noise model in the evolutionary search stage of QuantumNAS. Similarly, the \texttt{ibm\_kingston} noise model is used to generate the candidate circuits in the first stage of Élivágar and to compute Clifford noise resilience in the second stage. In contrast, the QuProFS work reports classification results only on the MNIST-2 dataset and not on MNIST-10; therefore, we directly use the MNIST-2 results reported in the QuProFS paper for comparative analysis.

\subsection{Ablation study}

\begin{table}[!b]
\centering
\caption{Performance comparison of DAQC based on \texttt{ibm\_cleveland} and \texttt{ibm\_kingston} quantum hardware without any error mitigation on PneumoniaMNIST-2 dataset using all test samples.}\vspace{8pt}
\label{table1}
\begin{tabular}{|cccccc|}      
\hline
\multicolumn{1}{|c|}{\textbf{Experiments}}                                                                                                     & \multicolumn{1}{c|}{\textbf{AUC}} & \multicolumn{1}{c|}{\textbf{ACC}} & \multicolumn{1}{c|}{\textbf{Specificity}} & \multicolumn{1}{c|}{\textbf{Sensitivity}} & \textbf{F1-score} \\ \hline
\multicolumn{1}{|c|}{\begin{tabular}[c]{@{}c@{}}DAQC, Noiseless \end{tabular}}                               & \multicolumn{1}{c|}{0.9427}       & \multicolumn{1}{c|}{0.8718}       & \multicolumn{1}{c|}{0.7222}               & \multicolumn{1}{c|}{0.9615}               & 0.9036            \\ \hline
\multicolumn{1}{|c|}{\begin{tabular}[c]{@{}c@{}}DAQC, \texttt{ibm\_cleveland}\end{tabular}} & \multicolumn{1}{c|}{0.5122}       & \multicolumn{1}{c|}{0.625}        & \multicolumn{1}{c|}{0}                    & \multicolumn{1}{c|}{1}                    & 0.7692            \\ \hline
\multicolumn{1}{|c|}{\begin{tabular}[c]{@{}c@{}}DAQC, \texttt{ibm\_kingston} \end{tabular}}  & \multicolumn{1}{c|}{0.9361}       & \multicolumn{1}{c|}{0.8381}       & \multicolumn{1}{c|}{0.6111}               & \multicolumn{1}{c|}{0.9744}               & 0.8827            \\ \hline
\end{tabular}
\end{table}

In this subsection, we performed ablation studies to evaluate our DAQC under different settings: i) backend comparison, ii) evaluation under different error suppression/mitigation techniques compatible with SamplerV2 and EstimatorV2 primitives, and iii) effect of entangling layer depth. Table \ref{table1} compares DAQC executed without error suppression/mitigation on two different IBM backends including \texttt{ibm\_cleveland} (127 qubit Eagle R3 processor) and \texttt{ibm\_kingston} (156 qubit Heron R2 processor). We observe significant improvement in the performance using the newer generation Heron processor as compared to the Eagle processor. On \texttt{ibm\_kingston} the performance is within a small margin of the noiseless simulator (AUC 0.9361 vs. 0.9427; ACC 0.8381 vs. 0.8718), whereas \texttt{ibm\_cleveland} exhibits significant degradation (AUC 0.5122; ACC 0.625) with zero specificity and saturated sensitivity as shown in Table \ref{table1}.

\begin{table}[!h]
\centering
\caption{Ablation of different error suppression and mitigation techniques of SamplerV2 and EstimatorV2 primitive based on PneumoniaMNIST-2 dataset.}\vspace{8pt}
\label{table2}
\begin{tabular}{|c|c|c|c|c|c|c|}
\hline
\textbf{Primitive}         & \textbf{Experiments}                                                                                                    & \textbf{AUC} & \textbf{ACC} & \textbf{Specificity} & \textbf{Sensitivity} & \textbf{F1-score} \\ \hline
\multirow{5}{*}{Sampler}   & \begin{tabular}[c]{@{}c@{}}DAQC, Noiseless \end{tabular}                              & 0.9427       & 0.8718       & 0.7222               & 0.9615               & 0.9036            \\ \cline{2-7} 
                           & \begin{tabular}[c]{@{}c@{}}DAQC, \texttt{ibm\_kingston} \\ no error mitigation\\ \end{tabular} & 0.9361       & 0.8381       & 0.6111               & 0.9744               & 0.8827            \\ \cline{2-7} 
                           & \begin{tabular}[c]{@{}c@{}}DAQC, \texttt{ibm\_kingston} \\ with DD+Twir+M3 \\ \end{tabular}          & 0.944        & 0.8077       & 0.5085               & 0.9872               & 0.8652            \\ \hline
\multirow{3}{*}{Estimator} & \begin{tabular}[c]{@{}c@{}}DAQC, Noiseless \\ \end{tabular}                              & 0.9425       & 0.8702       & 0.7051               & 0.9692               & 0.9032            \\ \cline{2-7} 
                           & \begin{tabular}[c]{@{}c@{}}DAQC, \texttt{ibm\_kingston} \\ with DD+TREX+Twir+ZNE \\ \end{tabular}    & 0.9391       & 0.86         & 0.6575               & 0.9764               & 0.8986            \\ \hline
\end{tabular}
\end{table}

Our previous works \cite{singh2025benchmarkingmedmnistdatasetreal, jin2025qfgnquantumapproachhighfidelity} show the effect of different error mitigation/suppression techniques compatible with SamplerV2 and EstimatorV2. We observed in \cite{singh2025benchmarkingmedmnistdatasetreal} that the combination of DD+Twir+M3 outperforms the other choices including DD+Twir and M3 mitigation in the case of Sampler. Similarly, we seen in \cite{jin2025qfgnquantumapproachhighfidelity} that the combination of DD+TREX+Twir+ZNE is best among the other choices including DD only, DD+TREX, DD+TREX+Twir, and DD+TREX+Twir+ZNE+PEC, in the case of Estimator. Motivated by our previous work, we evaluated DAQC on \texttt{ibm\_kingston} using DD+Twir+M3 with Sampler and DD+TREX+Twir+ZNE with Estimator on the PneumoniaMNIST-2 dataset in Table \ref{table2}. For the sampler experiment using DD+Twir+M3, we used all the testing samples for the evaluation on real hardware. However, in the case of Estimator using DD+TREX+Twir+ZNE, we restrict the hardware run to 200 test samples because ZNE requires repeated noise-scaled executions, making full-test evaluation time consuming. On Sampler, enabling DD+Twir+M3 preserves AUC and sensitivity metrics relative to noiseless i.e., AUC (0.944 vs. 0.9427) and sensitivity (0.9872 vs. 9615) but the other metrics including accuracy, specificity, and F1-score drops from 0.8718 (noiseless) to 0.8077, 0.7222 to 0.5085, and 0.9036 to 0.8652, respectively. On Estimator using DD+TREX+Twir+ZNE, DAQC attains an AUC of 0.9391 and an ACC of 0.86 with balanced specificity/sensitivity (0.6575/0.9764) and a F1-score of 0.8986, closely tracking the noiseless results as shown in Table \ref{table2}. Based on this, we decided to employ Estimator with DD+TREX+Twir+ZNE in our framework to evaluate DAQC in all experiments. 

\begin{table}[!h]
\centering
\caption{Performance analysis of DAQC with different number of entangling layers based on PneumoniaMNIST-2 dataset.}\vspace{8pt}
\label{table3}
\begin{tabular}{|c|c|c|c|c|c|}
\hline
\textbf{\begin{tabular}[c]{@{}c@{}}Number of entangling layers in DAQC\\ (each layer has 16 ECR gates)\end{tabular}} & \textbf{AUC}    & \textbf{ACC}    & \textbf{Specificity} & \textbf{Sensitivity} & \textbf{F1-score} \\ \hline
2                                                                                                                    & 0.9257          & 0.8478          & 0.6496               & 0.9667               & 0.8881            \\ \hline
3                                                                                                                    & 0.9357          & 0.8365          & 0.5855               & \textbf{0.9872}      & 0.883             \\ \hline
\textbf{4}                                                                                                           & \textbf{0.9425} & \textbf{0.8702} & \textbf{0.7051}      & 0.9692               & \textbf{0.9032}   \\ \hline
5                                                                                                                    & 0.9323          & 0.8558          & 0.688                & 0.9564               & 0.8923            \\ \hline
6                                                                                                                    & 0.9372          & 0.851           & 0.6709               & 0.959                & 0.8894            \\ \hline
8                                                                                                                    & 0.9379          & 0.8446          & 0.6197               & 0.9795               & 0.8873            \\ \hline
12                                                                                                                   & 0.937           & 0.8253          & 0.5726               & 0.9769               & 0.8749            \\ \hline
16                                                                                                                   & 0.8792          & 0.7885          & 0.4872               & 0.9692               & 0.8514            \\ \hline
\end{tabular}
\end{table}

Table \ref{table3} shows the effect of entangling depth on the performance of DAQC in noiseless settings. Varying the number of ECR layers (each layer = 16 ECR gates in a ring) reveals a clear optimal performance at 4 layers (i.e., total 64 ECR gates): AUC 0.9425, ACC 0.8702, F1 0.9032, with the best specificity among the different choices. Fewer layers (2–3) under-entangle and leave some spatial correlations under-modeled; beyond 4, performance steadily declines (e.g., 8, 12, 16 layers). A cautious interpretation is twofold: (i) optimization/expressivity balance—adding too many layers can make the optimization landscape harder and lead to overfitting, and (ii) hardware alignment—more entangling layers translate into larger two-qubit depth after transpilation and two-qubit gates are the noisiest on current devices. In short, DAQC benefits from periodic but sparse entanglement: enough ECR to couple neighboring patches and expand the receptive field, but not so many layers that training becomes brittle or hardware noise dominates.

\subsection{Performance comparison of DAQC with classical baselines}

\begin{table}[!b]
\centering
\caption{Performance comparison between classical baselines and DAQC on simulator and real \texttt{ibm\_kingston} quantum hardware based on MNIST-2, MNIST-4, and MNIST-10 datasets. All the classical baselines leverage the full spatial information of input images of size 28$\times$28 but our DAQC uses input size of 16$\times$16 due to current limitations of quantum hardware.}\vspace{8pt}
\label{table4}
\begin{tabular}{|ccccccc|}
\hline
\multicolumn{7}{|c|}{\textbf{MNIST-2}}                                                                                                                                                                                                                                                                                                                                                                                                                                                                \\ \hline
\multicolumn{1}{|c|}{\textbf{Models}}                                                                                           & \multicolumn{1}{c|}{\textbf{\begin{tabular}[c]{@{}c@{}}Trainable \\ parameters\end{tabular}}} & \multicolumn{1}{c|}{\textbf{AUC}}                   & \multicolumn{1}{c|}{\textbf{ACC}}                   & \multicolumn{1}{c|}{\textbf{Specificity}}           & \multicolumn{1}{c|}{\textbf{Sensitivity}}           & \textbf{F1-score}              \\ \hline
\multicolumn{1}{|c|}{ResNet18}                                                                                                               & \multicolumn{1}{c|}{11,169,858}                                                               & \multicolumn{1}{c|}{\cellcolor[HTML]{FFFFFF}1}      & \multicolumn{1}{c|}{\cellcolor[HTML]{FFFFFF}1}      & \multicolumn{1}{c|}{1}                              & \multicolumn{1}{c|}{1}                              & 1                              \\ \hline
\multicolumn{1}{|c|}{ResNet50}                                                                                                               & \multicolumn{1}{c|}{23,504,450}                                                               & \multicolumn{1}{c|}{\cellcolor[HTML]{FFFFFF}1}      & \multicolumn{1}{c|}{\cellcolor[HTML]{FFFFFF}1}      & \multicolumn{1}{c|}{\cellcolor[HTML]{FFFFFF}1}      & \multicolumn{1}{c|}{\cellcolor[HTML]{FFFFFF}1}      & \cellcolor[HTML]{FFFFFF}1      \\ \hline
\multicolumn{1}{|c|}{DenseNet121}                                                                                                            & \multicolumn{1}{c|}{6,948,226}                                                                  & \multicolumn{1}{c|}{\cellcolor[HTML]{FFFFFF}1}      & \multicolumn{1}{c|}{\cellcolor[HTML]{FFFFFF}1}      & \multicolumn{1}{c|}{\cellcolor[HTML]{FFFFFF}1}      & \multicolumn{1}{c|}{\cellcolor[HTML]{FFFFFF}1}      & \cellcolor[HTML]{FFFFFF}1      \\ \hline
\multicolumn{1}{|c|}{EfficientB0}                                                                                                            & \multicolumn{1}{c|}{4,010,110}                                                                  & \multicolumn{1}{c|}{\cellcolor[HTML]{FFFFFF}1}      & \multicolumn{1}{c|}{\cellcolor[HTML]{FFFFFF}1}      & \multicolumn{1}{c|}{\cellcolor[HTML]{FFFFFF}1}      & \multicolumn{1}{c|}{\cellcolor[HTML]{FFFFFF}1}      & \cellcolor[HTML]{FFFFFF}1      \\ \hline
\multicolumn{1}{|c|}{\begin{tabular}[c]{@{}c@{}}DAQC, Noiseless \end{tabular}}                            & \multicolumn{1}{c|}{546}                                                                      & \multicolumn{1}{c|}{0.9994}                         & \multicolumn{1}{c|}{0.9957}                         & \multicolumn{1}{c|}{0.998}                          & \multicolumn{1}{c|}{0.9938}                         & 0.996                          \\ \hline
\multicolumn{1}{|c|}{\begin{tabular}[c]{@{}c@{}}DAQC, \texttt{ibm\_kingston} \\ with DD+TREX+Twir+ZNE \end{tabular}} & \multicolumn{1}{c|}{546}                                                                      & \multicolumn{1}{c|}{0.9998}                         & \multicolumn{1}{c|}{0.985}                          & \multicolumn{1}{c|}{0.9899}                         & \multicolumn{1}{c|}{0.9802}                         & 0.9851                         \\ \hline 
\multicolumn{7}{|c|}{\textbf{MNIST-4}}                                                                                                                                                                                                                                                                                                                                                                                                                                                                \\ \hline
\multicolumn{1}{|c|}{\textbf{Models}}                                                                                           & \multicolumn{1}{c|}{\textbf{\begin{tabular}[c]{@{}c@{}}Trainable \\ parameters\end{tabular}}} & \multicolumn{1}{c|}{\textbf{AUC}}                   & \multicolumn{1}{c|}{\textbf{ACC}}                   & \multicolumn{1}{c|}{\textbf{Specificity}}           & \multicolumn{1}{c|}{\textbf{Sensitivity}}           & \textbf{F1-score}              \\ \hline
\multicolumn{1}{|c|}{ResNet18}                                                                                                               & \multicolumn{1}{c|}{11,170,884}                                                                 & \multicolumn{1}{c|}{\cellcolor[HTML]{FFFFFF}1}      & \multicolumn{1}{c|}{\cellcolor[HTML]{FFFFFF}0.9993} & \multicolumn{1}{c|}{0.9998}                         & \multicolumn{1}{c|}{0.9992}                         & 0.9992                         \\ \hline
\multicolumn{1}{|c|}{ResNet50}                                                                                                               & \multicolumn{1}{c|}{23,508,548}                                                               & \multicolumn{1}{c|}{\cellcolor[HTML]{FFFFFF}1}      & \multicolumn{1}{c|}{\cellcolor[HTML]{FFFFFF}0.9993} & \multicolumn{1}{c|}{0.9998}                         & \multicolumn{1}{c|}{0.9992}                         & 0.9992                         \\ \hline
\multicolumn{1}{|c|}{DenseNet121}                                                                                                            & \multicolumn{1}{c|}{6,950,276}                                                                  & \multicolumn{1}{c|}{\cellcolor[HTML]{FFFFFF}1}      & \multicolumn{1}{c|}{\cellcolor[HTML]{FFFFFF}0.999}  & \multicolumn{1}{c|}{\cellcolor[HTML]{FFFFFF}0.9997} & \multicolumn{1}{c|}{\cellcolor[HTML]{FFFFFF}0.999}  & \cellcolor[HTML]{FFFFFF}0.999  \\ \hline
\multicolumn{1}{|c|}{EfficientB0}                                                                                                            & \multicolumn{1}{c|}{4,012,672}                                                                  & \multicolumn{1}{c|}{1}                              & \multicolumn{1}{c|}{0.9995}                         & \multicolumn{1}{c|}{0.9998}                         & \multicolumn{1}{c|}{0.9995}                         & 0.9995                         \\ \hline
\multicolumn{1}{|c|}{\begin{tabular}[c]{@{}c@{}}DAQC, Noiseless \end{tabular}}                            & \multicolumn{1}{c|}{580}                                                                      & \multicolumn{1}{c|}{0.9905}                         & \multicolumn{1}{c|}{0.9329}                         & \multicolumn{1}{c|}{0.9778}                         & \multicolumn{1}{c|}{0.9309}                         & 0.9308                         \\ \hline
\multicolumn{1}{|c|}{\begin{tabular}[c]{@{}c@{}}DAQC, \texttt{ibm\_kingston} \\ with DD+TREX+Twir+ZNE \end{tabular}} & \multicolumn{1}{c|}{580}                                                                      & \multicolumn{1}{c|}{0.9864}                         & \multicolumn{1}{c|}{0.905}                          & \multicolumn{1}{c|}{0.9689}                         & \multicolumn{1}{c|}{0.8966}                         & 0.8966                         \\ \hline
\multicolumn{7}{|c|}{\textbf{MNIST-10}}                                                                                                                                                                                                                                                                                                                                                                                                                                                               \\ \hline
\multicolumn{1}{|c|}{\textbf{Models}}                                                                                           & \multicolumn{1}{c|}{\textbf{\begin{tabular}[c]{@{}c@{}}Trainable \\ parameters\end{tabular}}} & \multicolumn{1}{c|}{\textbf{AUC}}                   & \multicolumn{1}{c|}{\textbf{ACC}}                   & \multicolumn{1}{c|}{\textbf{Specificity}}           & \multicolumn{1}{c|}{\textbf{Sensitivity}}           & \textbf{F1-score}              \\ \hline
\multicolumn{1}{|c|}{ResNet18}                                                                                                               & \multicolumn{1}{c|}{11,173,962}                                                                 & \multicolumn{1}{c|}{\cellcolor[HTML]{FFFFFF}1}      & \multicolumn{1}{c|}{\cellcolor[HTML]{FFFFFF}0.9955} & \multicolumn{1}{c|}{0.9995}                         & \multicolumn{1}{c|}{0.9954}                         & 0.9955                         \\ \hline
\multicolumn{1}{|c|}{ResNet50}                                                                                                               & \multicolumn{1}{c|}{23,520,842}                                                               & \multicolumn{1}{c|}{\cellcolor[HTML]{FFFFFF}0.9999} & \multicolumn{1}{c|}{\cellcolor[HTML]{FFFFFF}0.9951} & \multicolumn{1}{c|}{0.9995}                         & \multicolumn{1}{c|}{0.9951}                         & 0.9951                         \\ \hline
\multicolumn{1}{|c|}{DenseNet121}                                                                                                            & \multicolumn{1}{c|}{6,956,426}                                                                  & \multicolumn{1}{c|}{\cellcolor[HTML]{FFFFFF}0.9999} & \multicolumn{1}{c|}{\cellcolor[HTML]{FFFFFF}0.9954} & \multicolumn{1}{c|}{\cellcolor[HTML]{FFFFFF}0.9995} & \multicolumn{1}{c|}{\cellcolor[HTML]{FFFFFF}0.9953} & \cellcolor[HTML]{FFFFFF}0.9954 \\ \hline
\multicolumn{1}{|c|}{EfficientB0}                                                                                                            & \multicolumn{1}{c|}{4,020,358}                                                                  & \multicolumn{1}{c|}{0.9999}                         & \multicolumn{1}{c|}{0.9941}                         & \multicolumn{1}{c|}{0.9993}                         & \multicolumn{1}{c|}{0.994}                          & 0.9941                         \\ \hline
\multicolumn{1}{|c|}{\begin{tabular}[c]{@{}c@{}}DAQC, Noiseless \end{tabular}}                            & \multicolumn{1}{c|}{682}                                                                      & \multicolumn{1}{c|}{0.9589}                         & \multicolumn{1}{c|}{0.7662}                         & \multicolumn{1}{c|}{0.9741}                         & \multicolumn{1}{c|}{0.7627}                         & 0.7617                         \\ \hline
\multicolumn{1}{|c|}{\begin{tabular}[c]{@{}c@{}}DAQC, \texttt{ibm\_kingston} \\ with DD+TREX+Twir+ZNE \end{tabular}} & \multicolumn{1}{c|}{682}                                                                      & \multicolumn{1}{c|}{0.9476}                         & \multicolumn{1}{c|}{0.73}                           & \multicolumn{1}{c|}{0.97}                           & \multicolumn{1}{c|}{0.7238}                         & 0.7183                         \\ \hline
\end{tabular}
\end{table}

The classical baselines trained on full 28$\times$28 images with millions of parameters, unsurprisingly saturate almost all metrics on MNIST-2, MNIST-4, and MNIST-10. Interestingly, our DAQC operating on $16\times16$ inputs with only a few hundred parameters, gets very close to the classical baselines scores in the case of MNIST-2 and MNIST-4, and remains competitive even on MNIST-10. On MNIST-2, DAQC attains near-saturated noiseless results (AUC 0.9994, ACC 0.9957, specificity 0.998, sensitivity 0.9938, F1 0.996) despite using 546 parameters versus 4–24M for the classical baselines as shown in Table \ref{table4}. This supports the claim that structured encoding plus sparse, locality-aware entanglement can extract competitive decision signals at dramatically lower model complexity. The hardware run on \texttt{ibm\_kingston} with DD+TREX+Twir+ZNE preserves this near-saturation (AUC 0.9998, ACC 0.985, specificity 0.9899, sensitivity 0.9802, F1 0.9851) for MNIST-2, showing that for binary classification the circuit depth and mitigation are well matched to device noise and the decision boundary stays stable. MNIST-4 tells a similar story, with the classical baselines still hover at the ceiling (AUC 1 and ACC/F1 $\approx$ 0.999) but DAQC remains strong with AUC 0.9905, ACC 0.9329, F1 0.9308, and a notably high specificity of 0.9778. This means DAQC is not only classifying well but also rejecting wrong classes reliably. When executed on hardware, DAQC keeps a high AUC (0.9864) and only loses about 3\% absolute accuracy (to 0.905) from the noiseless runs as shown in Table \ref{table4}. The corresponding specificity/sensitivity pair (0.9689/0.8966) indicates a small recall drop on the real device, which is typical for multi-class runs under noise and could be reduced with simple calibration or class-balanced objectives. Importantly, the slight change in parameter count across MNIST-2/4/10 comes only from the final linear layer that maps expectation values to the required number of logits; the circuit itself is the same. MNIST-10 makes the capacity and resolution limits most visible. Classical baselines still deliver saturated AUC and ACC/F1 $\approx$ 0.994-0.995. DAQC, with 16×16 inputs and 682 parameters, reaches AUC 0.9589 and ACC/F1 around 0.76 in noiseless simulation. The hardware-mitigated run (AUC 0.9476, ACC 0.73, F1 0.7183) is comparable to the noiseless results, with only about a 3–4\% absolute drop in ACC/F1. This results reveal two aspects: (i) ranking quality of DAQC remains strong (high AUC), implying that the model separates class scores reasonably well; (ii) but decision calibration for 10-way classification is more difficult using only 16$\times$16 input features with a very small parameter count (682), producing more confusions at the argmax stage (hence lower ACC/F1) than the classical baselines. This pattern is expected because moving from binary/4-class to 10-class classification typically requires greater model capacity. Therefore, we anticipate that further scaling the DAQC design will elevate the accuracy-based metrics.

Looking across all three MNIST settings, DAQC tends to keep specificity high, while sensitivity decreases as the task becomes harder and as we move from simulator to hardware. This makes intuitive sense: the architecture’s sparse, locality-aware entanglement and compact parameterization capture the dominant, easy-to-reject patterns (hence high specificity) but may under-encode rarer or subtler class cues under noise, especially at 10-class granularity. The gap between simulation and hardware remains small on MNIST-2/4 and manageable on MNIST-10, underscoring that the mitigation stack (DD+TREX+Twir+ZNE) preserves a large share of simulated performance on a modern Heron-class backend. Overall, Table 4 supports the main claim: with a single circuit template, lower input resolution, and three orders of magnitude fewer parameters than classical baselines, DAQC can match them on the binary task, stay close on the 4-class task, and retain strong ranking quality on the 10-class task—leaving clear headroom for future gains via slightly richer feature maps or modest circuit scaling. Moreover, we use the same circuit template (same qubit count and interleaving schedule) across all three datasets and obtain consistently high AUC and solid accuracy, which supports the generalization capability of the DAQC rather than dataset-specific tuning.

\begin{table}[!ht]
\centering
\caption{Performance comparison between classical baselines and DAQC on simulator and real \texttt{ibm\_kingston} quantum hardware based on FashionMNIST-2, FashionMNIST-4, and FashionMNIST-10 datasets. All the classical baselines leverage the full spatial information of input images of size 28$\times$28 but our DAQC uses input size of 16$\times$16 due to current limitations of quantum hardware.}\vspace{8pt}
\label{table5}
\begin{tabular}{|ccccccc|}
\hline
\multicolumn{7}{|c|}{\textbf{FashionMNIST-2}}                                                                                                                                                                                                                                                                                                                                                                                                            \\ \hline
\multicolumn{1}{|c|}{\textbf{Models}}                                                                                                      & \multicolumn{1}{c|}{\textbf{\begin{tabular}[c]{@{}c@{}}Trainable \\ parameters\end{tabular}}} & \multicolumn{1}{c|}{\textbf{AUC}}                   & \multicolumn{1}{c|}{\textbf{ACC}}                   & \multicolumn{1}{c|}{\textbf{Specificity}}           & \multicolumn{1}{c|}{\textbf{Sensitivity}}           & \textbf{F1-score}              \\ \hline
\multicolumn{1}{|c|}{ResNet18}                                                                                                             & \multicolumn{1}{c|}{11,169,858}                    & \multicolumn{1}{c|}{\cellcolor[HTML]{FFFFFF}1}      & \multicolumn{1}{c|}{\cellcolor[HTML]{FFFFFF}0.996}  & \multicolumn{1}{c|}{0.999}                          & \multicolumn{1}{c|}{0.993}                          & 0.996                          \\ \hline
\multicolumn{1}{|c|}{ResNet50}                                                                                                             & \multicolumn{1}{c|}{23,504,450}                    & \multicolumn{1}{c|}{\cellcolor[HTML]{FFFFFF}0.9994} & \multicolumn{1}{c|}{\cellcolor[HTML]{FFFFFF}0.9955} & \multicolumn{1}{c|}{\cellcolor[HTML]{FFFFFF}0.998}  & \multicolumn{1}{c|}{\cellcolor[HTML]{FFFFFF}0.993}  & \cellcolor[HTML]{FFFFFF}0.9955 \\ \hline
\multicolumn{1}{|c|}{DenseNet121}                                                                                                          & \multicolumn{1}{c|}{6,948,226}                       & \multicolumn{1}{c|}{\cellcolor[HTML]{FFFFFF}1}      & \multicolumn{1}{c|}{\cellcolor[HTML]{FFFFFF}0.998}  & \multicolumn{1}{c|}{\cellcolor[HTML]{FFFFFF}0.997}  & \multicolumn{1}{c|}{\cellcolor[HTML]{FFFFFF}0.999}  & \cellcolor[HTML]{FFFFFF}0.998  \\ \hline
\multicolumn{1}{|c|}{EfficientB0}                                                                                                          & \multicolumn{1}{c|}{4,010,110}                       & \multicolumn{1}{c|}{\cellcolor[HTML]{FFFFFF}0.999}  & \multicolumn{1}{c|}{\cellcolor[HTML]{FFFFFF}0.996}  & \multicolumn{1}{c|}{\cellcolor[HTML]{FFFFFF}1}      & \multicolumn{1}{c|}{\cellcolor[HTML]{FFFFFF}0.992}  & \cellcolor[HTML]{FFFFFF}0.996  \\ \hline
\multicolumn{1}{|c|}{\begin{tabular}[c]{@{}c@{}}DAQC, Noiseless \end{tabular}}                           & \multicolumn{1}{c|}{\cellcolor[HTML]{FFFFFF}546}   & \multicolumn{1}{c|}{\cellcolor[HTML]{FFFFFF}0.9956} & \multicolumn{1}{c|}{\cellcolor[HTML]{FFFFFF}0.974}  & \multicolumn{1}{c|}{\cellcolor[HTML]{FFFFFF}0.986}  & \multicolumn{1}{c|}{\cellcolor[HTML]{FFFFFF}0.962}  & \cellcolor[HTML]{FFFFFF}0.9737 \\ \hline
\multicolumn{1}{|c|}{\begin{tabular}[c]{@{}c@{}}DAQC, \texttt{ibm\_kingston} \\ with DD+TREX+Twir+ZNE \end{tabular}} & \multicolumn{1}{c|}{546}                           & \multicolumn{1}{c|}{0.9978}                         & \multicolumn{1}{c|}{0.99}                           & \multicolumn{1}{c|}{1}                              & \multicolumn{1}{c|}{0.9796}                         & 0.9897                         \\ \hline
\multicolumn{7}{|c|}{\textbf{FashionMNIST-4}}                                                                                                                                                                                                                                                                                                                                                                                                            \\ \hline
\multicolumn{1}{|c|}{\textbf{Models}}                                                                                                      & \multicolumn{1}{c|}{\textbf{\begin{tabular}[c]{@{}c@{}}Trainable \\ parameters\end{tabular}}} & \multicolumn{1}{c|}{\textbf{AUC}}                   & \multicolumn{1}{c|}{\textbf{ACC}}                   & \multicolumn{1}{c|}{\textbf{Specificity}}           & \multicolumn{1}{c|}{\textbf{Sensitivity}}           & \textbf{F1-score}              \\ \hline
\multicolumn{1}{|c|}{ResNet18}                                                                                                             & \multicolumn{1}{c|}{11,170,884}                      & \multicolumn{1}{c|}{\cellcolor[HTML]{FFFFFF}0.9998} & \multicolumn{1}{c|}{\cellcolor[HTML]{FFFFFF}0.9952} & \multicolumn{1}{c|}{0.9984}                         & \multicolumn{1}{c|}{0.9953}                         & 0.9953                         \\ \hline
\multicolumn{1}{|c|}{ResNet50}                                                                                                             & \multicolumn{1}{c|}{23,508,548}                    & \multicolumn{1}{c|}{\cellcolor[HTML]{FFFFFF}0.9999} & \multicolumn{1}{c|}{\cellcolor[HTML]{FFFFFF}0.995}  & \multicolumn{1}{c|}{0.9983}                         & \multicolumn{1}{c|}{0.995}                          & 0.995                          \\ \hline
\multicolumn{1}{|c|}{DenseNet121}                                                                                                          & \multicolumn{1}{c|}{6,950,276}                       & \multicolumn{1}{c|}{\cellcolor[HTML]{FFFFFF}0.9996} & \multicolumn{1}{c|}{\cellcolor[HTML]{FFFFFF}0.9968} & \multicolumn{1}{c|}{\cellcolor[HTML]{FFFFFF}0.9989} & \multicolumn{1}{c|}{\cellcolor[HTML]{FFFFFF}0.9968} & \cellcolor[HTML]{FFFFFF}0.9968 \\ \hline
\multicolumn{1}{|c|}{EfficientB0}                                                                                                          & \multicolumn{1}{c|}{4,012,672}                       & \multicolumn{1}{c|}{0.9993}                         & \multicolumn{1}{c|}{0.9942}                         & \multicolumn{1}{c|}{0.9981}                         & \multicolumn{1}{c|}{0.9942}                         & 0.9943                         \\ \hline
\multicolumn{1}{|c|}{\begin{tabular}[c]{@{}c@{}}DAQC, Noiseless \end{tabular}}                           & \multicolumn{1}{c|}{580}                           & \multicolumn{1}{c|}{0.9923}                         & \multicolumn{1}{c|}{0.9377}                         & \multicolumn{1}{c|}{0.9792}                         & \multicolumn{1}{c|}{0.9377}                         & 0.9379                         \\ \hline
\multicolumn{1}{|c|}{\begin{tabular}[c]{@{}c@{}}DAQC, \texttt{ibm\_kingston} \\ with DD+TREX+Twir+ZNE \end{tabular}} & \multicolumn{1}{c|}{580}                           & \multicolumn{1}{c|}{0.9889}                         & \multicolumn{1}{c|}{0.945}                          & \multicolumn{1}{c|}{0.9812}                         & \multicolumn{1}{c|}{0.9462}                         & 0.947                          \\ \hline
\multicolumn{7}{|c|}{\textbf{FashionMNIST-10}}                                                                                                                                                                                                                                                                                                                                                                                                           \\ \hline
\multicolumn{1}{|c|}{\textbf{Models}}                                                                                                      & \multicolumn{1}{c|}{\textbf{\begin{tabular}[c]{@{}c@{}}Trainable \\ parameters\end{tabular}}} & \multicolumn{1}{c|}{\textbf{AUC}}                   & \multicolumn{1}{c|}{\textbf{ACC}}                   & \multicolumn{1}{c|}{\textbf{Specificity}}           & \multicolumn{1}{c|}{\textbf{Sensitivity}}           & \textbf{F1-score}              \\ \hline
\multicolumn{1}{|c|}{ResNet18}                                                                                                             & \multicolumn{1}{c|}{11,173,962}                      & \multicolumn{1}{c|}{\cellcolor[HTML]{FFFFFF}0.992}  & \multicolumn{1}{c|}{\cellcolor[HTML]{FFFFFF}0.9262} & \multicolumn{1}{c|}{0.9918}                         & \multicolumn{1}{c|}{0.9262}                         & 0.9258                         \\ \hline
\multicolumn{1}{|c|}{ResNet50}                                                                                                             & \multicolumn{1}{c|}{23,520,842}                    & \multicolumn{1}{c|}{\cellcolor[HTML]{FFFFFF}0.9935} & \multicolumn{1}{c|}{\cellcolor[HTML]{FFFFFF}0.9295} & \multicolumn{1}{c|}{0.9922}                         & \multicolumn{1}{c|}{0.9295}                         & 0.929                          \\ \hline
\multicolumn{1}{|c|}{DenseNet121}                                                                                                          & \multicolumn{1}{c|}{6,956,426}                       & \multicolumn{1}{c|}{\cellcolor[HTML]{FFFFFF}0.9941} & \multicolumn{1}{c|}{\cellcolor[HTML]{FFFFFF}0.9386} & \multicolumn{1}{c|}{\cellcolor[HTML]{FFFFFF}0.9932} & \multicolumn{1}{c|}{\cellcolor[HTML]{FFFFFF}0.9386} & \cellcolor[HTML]{FFFFFF}0.9384 \\ \hline
\multicolumn{1}{|c|}{EfficientB0}                                                                                                          & \multicolumn{1}{c|}{4,020,358}                       & \multicolumn{1}{c|}{0.9916}                         & \multicolumn{1}{c|}{0.9228}                         & \multicolumn{1}{c|}{0.9914}                         & \multicolumn{1}{c|}{0.9228}                         & 0.9227                         \\ \hline
\multicolumn{1}{|c|}{\begin{tabular}[c]{@{}c@{}}DAQC, Noiseless \end{tabular}}                           & \multicolumn{1}{c|}{682}                           & \multicolumn{1}{c|}{0.9447}                         & \multicolumn{1}{c|}{0.6888}                         & \multicolumn{1}{c|}{0.9654}                         & \multicolumn{1}{c|}{0.6888}                         & 0.6851                         \\ \hline
\multicolumn{1}{|c|}{\begin{tabular}[c]{@{}c@{}}DAQC, \texttt{ibm\_kingston} \\ with DD+TREX+Twir+ZNE \end{tabular}} & \multicolumn{1}{c|}{682}                           & \multicolumn{1}{c|}{0.954}                          & \multicolumn{1}{c|}{0.71}                           & \multicolumn{1}{c|}{0.968}                          & \multicolumn{1}{c|}{0.7422}                         & 0.7111                         \\ \hline
\end{tabular}
\end{table}

On FashionMNIST, the situation is similar to MNIST: the classical baselines again set a high bar with full-resolution 28×28 inputs and millions of parameters, but the proposed DAQC—working with compressed 16×16 inputs and only hundreds of learnable weights remains competitive in the binary and 4-class settings and preserves strong ranking quality even at 10 classes. For FashionMNIST-2, DAQC nearly saturates every metric. The noiseless run gives AUC 0.9956, ACC 0.974, F1 0.9737, and a balanced specificity/sensitivity pair (0.986/0.962). The hardware run with DD+TREX+Twir+ZNE tracks noiseless results closely with AUC 0.9978, ACC 0.99, specificity 1.0, sensitivity 0.9796, and F1 0.9897, indicating that circuit depth plus mitigation is well matched to \texttt{ibm\_kingston} as shown in Table \ref{table5}. On FashionMNIST-4, the classical models still hover around perfect scores, but DAQC remains strong: in simulation runs it achieves AUC 0.9923, ACC 0.9377, F1 0.9379, and high specificity (0.9792) with only 580 parameters. The hardware run shows an interesting pattern: AUC dips slightly relative to the noiseless (0.9889 vs. 0.9923) but slightly better ACC and F1 (0.945 and 0.947) and still very good specificity/sensitivity (0.9812/0.9462). That kind of small reshuffling is reasonable when you evaluate on a finite subset and apply noise mitigation—scores get slightly re-ordered, but the model’s overall discrimination ability stays high. FashionMNIST-10 is, as expected, the hardest case. The classical baselines stays around 0.93–0.94 ACC/F1, while DAQC, limited to 16$\times$16 and 682 parameters, reaches AUC 0.9447 and ACC/F1 $\approx$ 0.69 in the noiseless simulation as shown in Table \ref{table5}. This shows that the model is still ranking classes well (high AUC) but loses more at the final 10-way decision step because of capacity and input compression similar to the MNIST-10. The hardware run with mitigation provides comparable results to the noiseless case but we noticed improvement in the decision metrics: ACC increases from 0.6888 to 0.71 and F1-score from 0.6851 to 0.7111, mainly because sensitivity improves (0.7422) while specificity stays high (0.968). In other words, the hardware plus mitigation pipeline shifts the model toward a more recall-friendly operating point. Across the three FashionMNIST partitions, DAQC exhibits a consistent specificity advantage with sensitivity gradually tightening as the task complexity rises. In binary and 4-class settings, sensitivity remains high and close to specificity; in the 10-class it becomes the limiting factor, reflecting tighter capacity and the information bottleneck from 16$\times$16 pooling.

\begin{table}[!ht]
\centering
\caption{Performance comparison between classical baselines and proposed DAQC on simulator and real \texttt{ibm\_kingston} quantum hardware based on PneumoniaMNIST-2 dataset. All the classical baselines leverage the full spatial information of input images of size 28$\times$28 but our DAQC uses input size of 16$\times$16 due to current limitations of quantum hardware.}\vspace{8pt}
\label{table6}
\begin{tabular}{|ccccccc|}
\hline
\multicolumn{7}{|c|}{\textbf{PneumoniaMNIST-2}}                                                                                                                                                                                                                                                                                                                                                                                                                   \\ \hline
\multicolumn{1}{|c|}{\textbf{Models}}                                                                                                      & \multicolumn{1}{c|}{\textbf{\begin{tabular}[c]{@{}c@{}}Trainable\\ parameters\end{tabular}}} & \multicolumn{1}{c|}{\textbf{AUC}}                   & \multicolumn{1}{c|}{\textbf{ACC}}                   & \multicolumn{1}{c|}{\textbf{Specificity}} & \multicolumn{1}{c|}{\textbf{Sensitivity}} & \textbf{F1-score} \\ \hline
\multicolumn{1}{|c|}{ResNet18}                                                                                                             & \multicolumn{1}{c|}{11,169,858}                                                              & \multicolumn{1}{c|}{\cellcolor[HTML]{FFFFFF}0.9329} & \multicolumn{1}{c|}{\cellcolor[HTML]{FFFFFF}0.8558} & \multicolumn{1}{c|}{0.6282}               & \multicolumn{1}{c|}{0.9923}               & 0.8958            \\ \hline
\multicolumn{1}{|c|}{ResNet50}                                                                                                             & \multicolumn{1}{c|}{23,504,450}                                                              & \multicolumn{1}{c|}{\cellcolor[HTML]{FFFFFF}0.9345} & \multicolumn{1}{c|}{\cellcolor[HTML]{FFFFFF}0.851}  & \multicolumn{1}{c|}{0.6111}               & \multicolumn{1}{c|}{0.9949}               & 0.893             \\ \hline
\multicolumn{1}{|c|}{DenseNet121}                                                                                                          & \multicolumn{1}{c|}{6,948,226}                                                               & \multicolumn{1}{c|}{0.9745}                         & \multicolumn{1}{c|}{0.8782}                         & \multicolumn{1}{c|}{0.6838}               & \multicolumn{1}{c|}{0.9949}               & 0.9108            \\ \hline
\multicolumn{1}{|c|}{EfficientB0}                                                                                                          & \multicolumn{1}{c|}{4,010,110}                                                               & \multicolumn{1}{c|}{0.9338}                         & \multicolumn{1}{c|}{0.8301}                         & \multicolumn{1}{c|}{0.5556}               & \multicolumn{1}{c|}{0.9949}               & 0.8798            \\ \hline
\multicolumn{1}{|c|}{\begin{tabular}[c]{@{}c@{}}DAQC, Noiseless \end{tabular}}                           & \multicolumn{1}{c|}{546}                                                                     & \multicolumn{1}{c|}{0.9425}                         & \multicolumn{1}{c|}{0.8702}                         & \multicolumn{1}{c|}{0.7051}               & \multicolumn{1}{c|}{0.9692}               & 0.9032            \\ \hline
\multicolumn{1}{|c|}{\begin{tabular}[c]{@{}c@{}}DAQC, \texttt{ibm\_kingston} \\ with DD+TREX+Twir+ZNE \end{tabular}} & \multicolumn{1}{c|}{546}                                                                     & \multicolumn{1}{c|}{0.9391}                         & \multicolumn{1}{c|}{0.86}                           & \multicolumn{1}{c|}{0.6575}               & \multicolumn{1}{c|}{0.9764}               & 0.8986            \\ \hline
\end{tabular}
\end{table}

For PneumoniaMNIST-2, the classical baselines again benefit from full 28×28 inputs and million number of parameters, but DAQC—restricted to 16$\times$16 inputs and only 546 trainable parameters—remains competitive and, in fact, gives a more balanced operating point. ResNet18, ResNet50, and EfficientB0 all reach reasonable AUC (0.9329–0.9345) and ACC (0.8301–0.8558), but they do so with a strong sensitivity–specificity imbalance: sensitivity is almost saturated ($\approx$ 0.995) while specificity is low ($\approx$ 0.56–0.63) as shown in Table \ref{table6}. DenseNet121 is the strongest classical model (AUC 0.9745, ACC 0.8782, F1 0.9108) and improves specificity to 0.6838, but it still shows the same recall-heavy, false-positive-prone behavior. DAQC’s noiseless performance (AUC 0.9425, ACC 0.8702, F1 0.9032) compares favorably to the classical baselines, and importantly it does so with a better operating balance: specificity 0.7051 and sensitivity 0.9692. Relative to the best performing DenseNet121 model among classical baselines, DAQC forfeits some AUC but actually improves specificity (0.7051 vs. 0.6838) while keeping sensitivity high. This suggests that the locality-aware encoding and sparse entanglement help the model to capture clinically relevant negatives more faithfully, reducing the rate of false alarms without collapsing recall. Executed on \texttt{ibm\_kingston} with DD+TREX+Twir+ZNE, DAQC preserves the overall profile with only modest drift: AUC 0.9391, ACC 0.86, F1 0.8986, sensitivity 0.9764 and specificity 0.6575 which is better than many classical baselines as shown in Table \ref{table6}. These are typical, small degradations under real-device noise and show that the ranking ability remains stable. Overall, DAQC reaches AUC $\approx$ 0.94 and F1 $\approx$ 0.90 with only 546 parameters, whereas the classical baselines use 4–24 million parameters, and even the best classical model does not clearly dominate DAQC on specificity. Given the documented class imbalance in PneumoniaMNIST-2, the fact that DAQC keeps sensitivity high without collapsing specificity suggests the circuit’s bias–variance trade-off is well tuned to the dataset’s skew and to the compressed 16$\times$16 representation.

\subsection{Performance comparison of DAQC with QCS baselines}

We also compared the performance of DAQC with exiting QCS methods on the MNIST-10 dataset in Table \ref{table7}. Due to limited access to quantum hardware, we report noiseless results for all evaluated QCS approaches. However, it is worth noting that hardware runs of these methods typically further degrade performance metrics because of noise. Table \ref{table7} highlights a sharp performance gap between DAQC and prior QCS baselines on MNIST-10. The methods QuantumSupernet and QuantumNAS deliver low AUCs ($\approx$ 0.54–0.55) and very low accuracies ($\approx$ 0.12–0.15), with very low F1-scores ($\approx$ 0.09–0.10). Élivágar improves substantially over QuantumSupernet and QuantumNAS with AUC 0.7673, ACC 0.3604, and F1 0.3184, yet still underperforms by a wide margin relative to DAQC. In contrast, DAQC reaches AUC 0.9589, ACC 0.7662, and F1 0.7617 in noiseless settings and maintains strong performance on real hardware (AUC 0.9476, ACC 0.73, F1 0.7183). These results indicate not only better hard decisions but also a markedly superior score ordering: DAQC separates classes far more cleanly than its QCS counterparts, which is crucial in a 10-class task where decision boundaries are inherently tighter. Moreover, on the MNIST-2 dataset, the recent QuProFS framework achieves noiseless and hardware accuracies of 0.99 and 0.92, respectively, whereas DAQC attains 0.9957 and 0.985. Thus, DAQC outperforms QuProFS by providing an accuracy improvement of 6.5\% on real hardware.

\begin{table}[!ht]
\centering
\caption{Comparative analysis of DAQC with the existing SOTA approaches based on MNIST-10 dataset.}\vspace{8pt}
\label{table7}
\begin{tabular}{|cccccc|}
\hline
\multicolumn{6}{|c|}{\textbf{MNIST-10}}                                                                                                                                                                                                                                                                  \\ \hline
\multicolumn{1}{|c|}{\textbf{Models}}                                                                          & \multicolumn{1}{c|}{\textbf{AUC}}    & \multicolumn{1}{c|}{\textbf{ACC}}    & \multicolumn{1}{c|}{\textbf{Specificity}} & \multicolumn{1}{c|}{\textbf{Sensitivity}} & \textbf{F1-score} \\ \hline
\multicolumn{1}{|c|}{QuantumSupernet}                                                                   & \multicolumn{1}{c|}{0.5409}          & \multicolumn{1}{c|}{0.1453}          & \multicolumn{1}{c|}{0.9045}               & \multicolumn{1}{c|}{0.1381}               & 0.1014            \\ \hline
\multicolumn{1}{|c|}{QuantumNAS}                                                                        & \multicolumn{1}{c|}{0.5491}          & \multicolumn{1}{c|}{0.1241}          & \multicolumn{1}{c|}{0.9027}               & \multicolumn{1}{c|}{0.1264}               & 0.0875            \\ \hline
\multicolumn{1}{|c|}{Elivagar}                                                                          & \multicolumn{1}{c|}{0.7673}          & \multicolumn{1}{c|}{0.3604}          & \multicolumn{1}{c|}{0.9288}               & \multicolumn{1}{c|}{0.3522}               & 0.3184            \\ \hline
\multicolumn{1}{|c|}{DAQC (Noiseless)}                                                                 & \multicolumn{1}{c|}{\textbf{0.9589}} & \multicolumn{1}{c|}{\textbf{0.7662}} & \multicolumn{1}{c|}{\textbf{0.9741}}      & \multicolumn{1}{c|}{\textbf{0.7627}}      & \textbf{0.7617}   \\ \hline
\multicolumn{1}{|c|}{\begin{tabular}[c]{@{}c@{}}DAQC, \texttt{ibm\_kingston} \\ with DD+TREX+Twir+ZNE\end{tabular}} & \multicolumn{1}{c|}{\textbf{0.9476}} & \multicolumn{1}{c|}{\textbf{0.73}}   & \multicolumn{1}{c|}{\textbf{0.97}}        & \multicolumn{1}{c|}{\textbf{0.7238}}      & \textbf{0.7183}   \\ \hline
\end{tabular}
\end{table}

This comparison is also fair (i) all methods are evaluated on MNIST-10, which prevents inflated results from an almost linearly separable binary task and (ii) to align with our hardware deployment, the \texttt{ibm\_kingston} noise model informs the search/evaluation stages of QuantumNAS and Élivágar, reducing the risk that DAQC’s advantage stems purely from a mismatch between searched circuits and deployment noise. Even under that fairness adjustment, DAQC leads decisively across all metrics. Moreover, the simulator-hardware gap for DAQC is modest (AUC drop $\approx$ 0.011, ACC drop $\approx$ 0.036, F1 drop $\approx$ 0.043), indicating that the circuit’s locality-aware entanglement and encoding translate robustly to real devices when paired with DD+TREX+Twir+ZNE. In practical terms, DAQC establishes a better Pareto point: high accuracy and F1, high specificity and usable sensitivity, and only modest degradation on hardware — which supports the argument that domain-aware, hardware-aligned circuit design can outperform unconstrained circuit search when moving to realistic multi-class problems.

\subsection{Barren plateau analysis of DAQC}

Training deep variational circuits is vulnerable to barren plateaus, where gradient magnitudes vanish exponentially in the number of qubits or depth, making optimization effectively impossible. Our DAQC design choices (discussed in Section 2) helps to tackle the barren plateau, and to validate this, we analyze the trainability of DAQC ansatz in two complementary ways: (i) McClean-style \cite{McClean_2018} random-initialization scaling on ideal statevector simulations and (ii) gradient statistics along the actual supervised training trajectory.

\subsubsection{Gradient variance at random initialization}

Building on standard barren plateau analyses~\cite{McClean_2018,Cerezo_2021}, we study two idealized cost functions for random instances of our DAQC ansatz, \(C_{global}(\boldsymbol\theta) = \big\langle Z^{\otimes n} \big\rangle\) and \(C_{local}(\boldsymbol\theta) = \big\langle Z_0 \big\rangle\), where \(C_{global}\) probes a fully global observable on all $n$ qubits and \(C_{local}\) depends on a single-qubit (1-local) operator. These costs are used purely as diagnostic probes of global versus local scaling. Our actual training objective is a cross-entropy loss built from the vector of single-qubit expectations \(\big(\langle Z_0\rangle,\ldots,\langle Z_{n-1}\rangle\big)\), and therefore
belongs to the same local-cost class as \(C_{local}\). For each configuration of interest (number of qubits, number of trainable parameters, and layer index for every trainable column depth), we generate an ensemble of random DAQC-style circuits as described in Section 2. For each configuration, we sample five such circuit instances. For a given circuit, we draw a single data-encoding vector $x \sim U[0,\pi]$ and 50 random weight vectors $\boldsymbol\theta \sim U(0,2\pi)$, evaluate the chosen cost $C(\boldsymbol\theta)$ exactly on a GPU-accelerated statevector simulator, and compute $\partial_{\theta_j} C$ for all trainable parameters using the parameter-shift rule. From these samples we estimate, for each parameter $j$, the gradient variance $Var[\partial_{\theta_j} C]$ and then aggregate across all parameters of all circuits to obtain the mean
gradient variance
\(
\frac{1}{D}\sum_{j=1}^{D} {Var}[\partial_{\theta_j} C]
\),
where $D$ is the total number of trainable parameters in the ensemble. Moreover, we compute a layer-wise mean variance by grouping
parameters according to their layer index to study the barren plateau effect along the depth of DAQC.

\begin{figure}[!ht]
  \centering
  \begin{subfigure}[t]{0.33\linewidth}
    \centering
    \includegraphics[width=\linewidth]{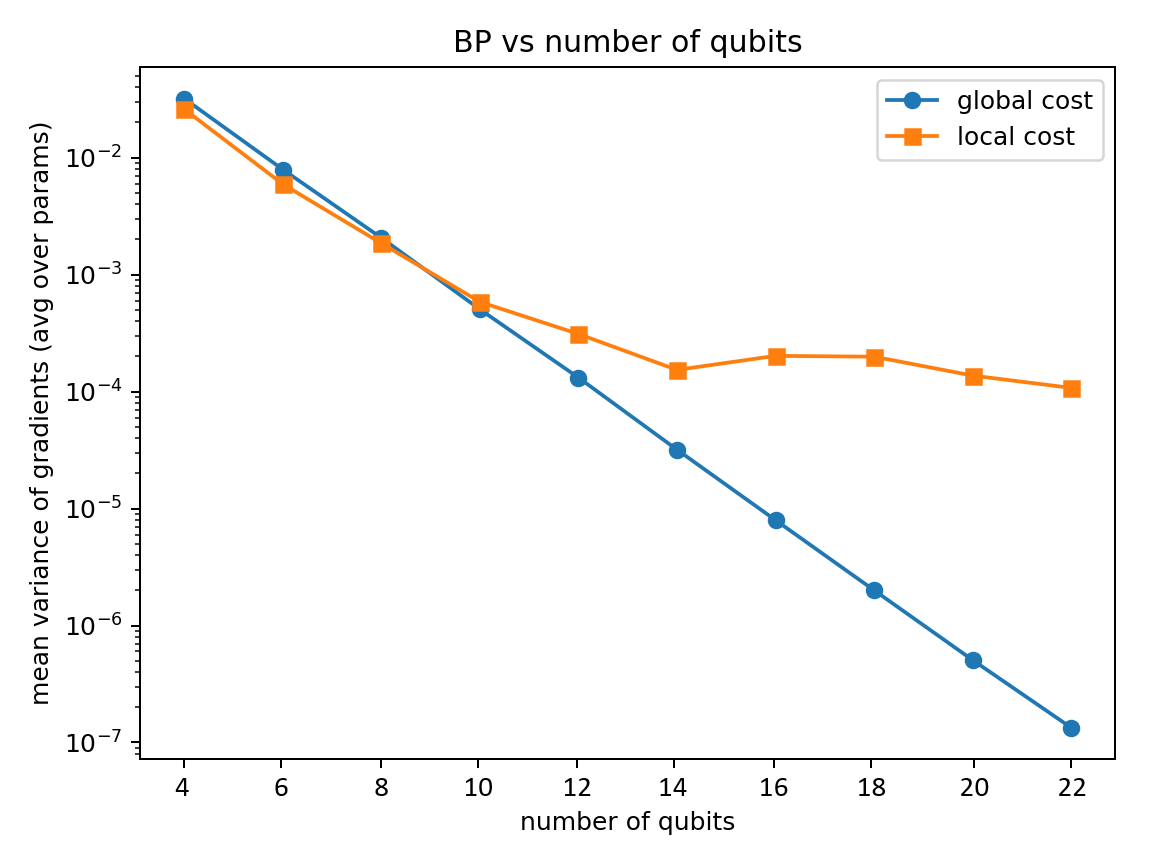}
    \caption{}
    \label{fig:daqc-a}
  \end{subfigure}\hfill
  \begin{subfigure}[t]{0.33\linewidth}
    \centering
    \includegraphics[width=\linewidth]{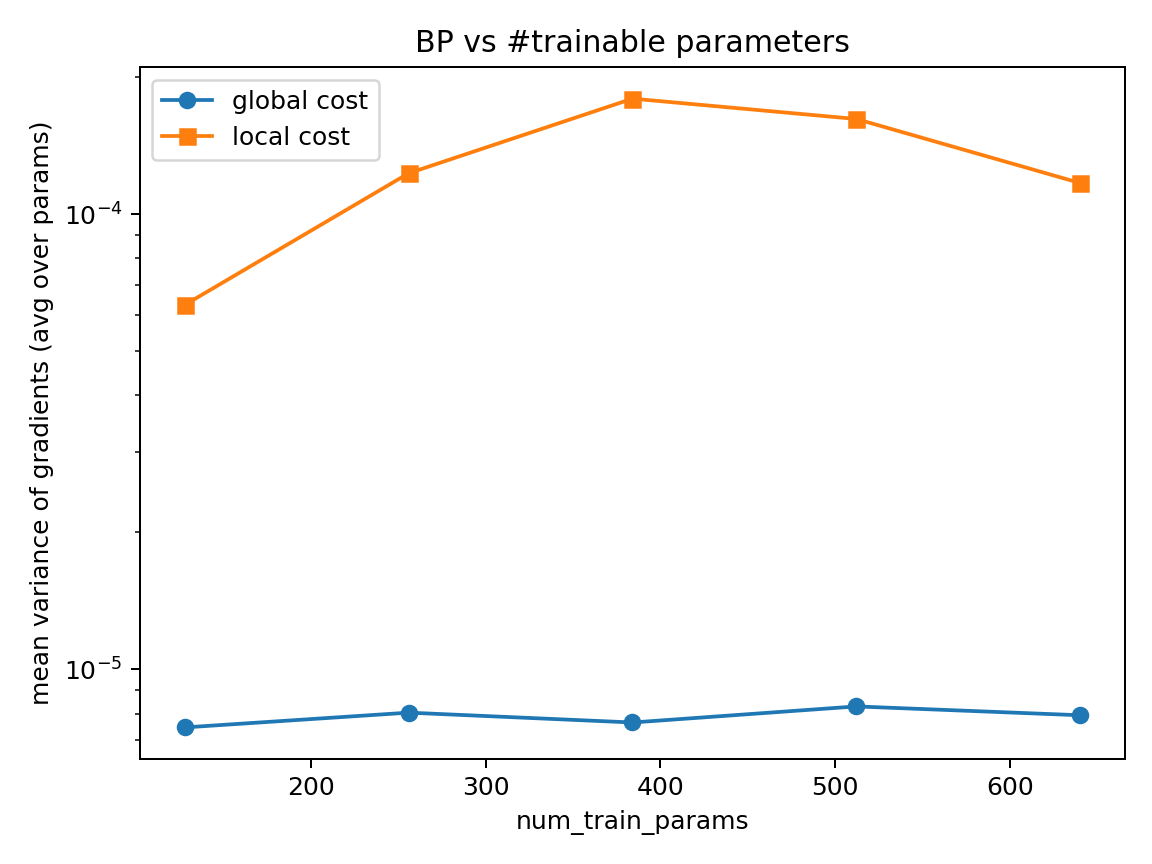}
    \caption{}
    \label{fig:daqc-b}
  \end{subfigure}\hfill
  \begin{subfigure}[t]{0.33\linewidth}
    \centering
    \includegraphics[width=\linewidth]{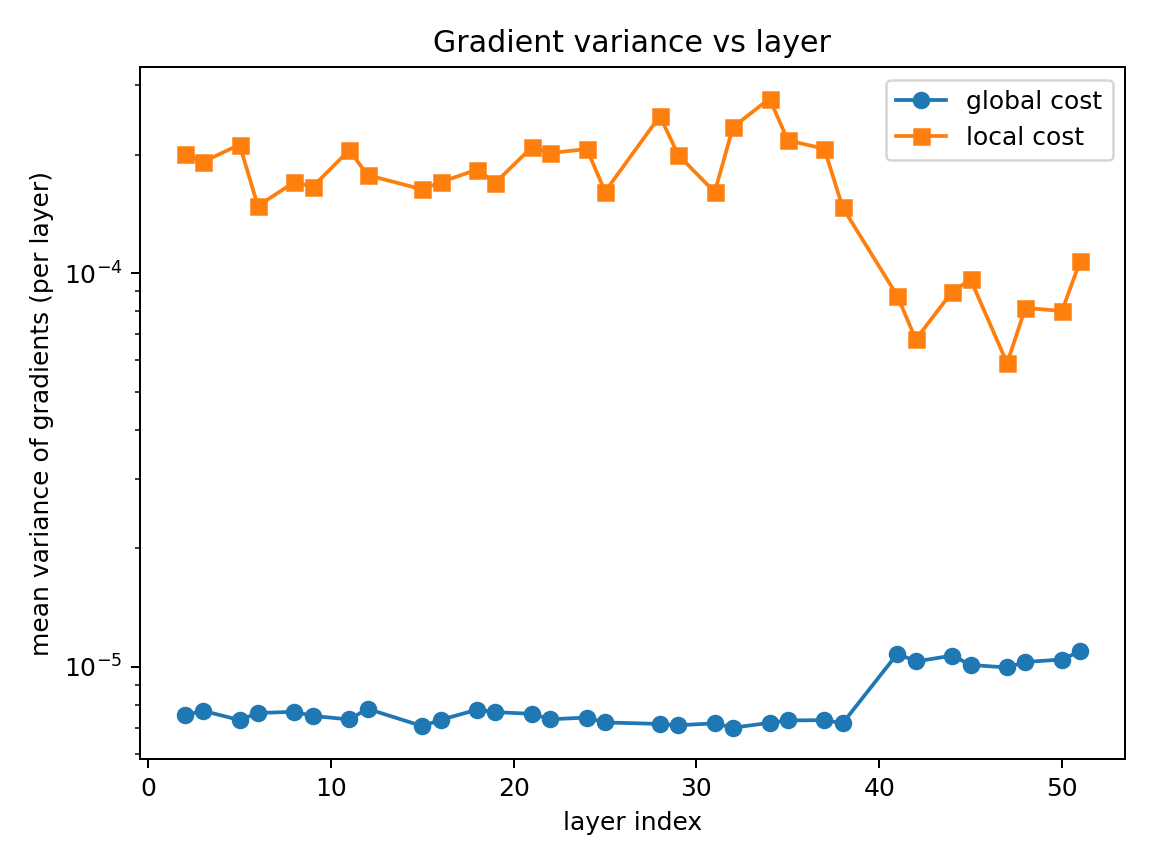}
    \caption{}
    \label{fig:daqc-c}
  \end{subfigure}
\caption{Barren plateau analysis for the DAQC ansatz in McClean-style. Each plot reports the mean gradient variance at random initialization, averaged over all trainable parameters and an ensemble of random circuits: (a) Scaling with the number of qubits, (b) Dependence on the number of trainable parameters, and (c) Layer-wise mean gradient variance.}
\label{figure6}
\end{figure}

Fig. \ref{figure6}a shows the mean gradient variance as a function of the number of qubits while keeping the per-qubit depth fixed. Concretely, we scale the total number of embedding and trainable gates linearly with the number of qubits so that depth-per-qubit template is preserved as we sweep $n \in \{4,6,8,10,12,14,16,18,20,22\}$. On a semilog scale, the variance
for the global cost $C_{global}$ decays almost linearly in $n$, dropping from $\sim 10^{-2}$ at $n=4$ to $\sim 10^{-7}$ at $n=22$, an essentially exponential barren-plateau scaling. In contrast, the variance for the local cost $C_{local}$ decreases initially but quickly saturates in the $10^{-4}$ range and remains between roughly one and three orders of magnitude larger than the global counterpart; by $n\approx 20$-$22$ the gap approaches three orders of magnitude. This behaviour is consistent with cost-function-dependent theory: locality of the cost strongly mitigates the worst exponential decay with system
size \cite{Cerezo_2021}. 

We also investigate the DAQC by varying the number of trainable parameters while holding the number of qubits and encoding gates fixed to 16 and 256, respectively in Fig. \ref{figure6}b. We sweep number of trainable parameters in $\{128,256,384,512,640\}$ while keeping the DAQC layout, entangling pattern, and other hyperparameters unchanged. For the global cost, the mean gradient variance remains in a narrow band around $\sim 10^{-5}$ across this entire range: at $n=16$ the gradients are already in the small regime dictated by the qubit-scaling behavior in Fig. \ref{figure6}a, and increasing trainable parameters does not induce exponential gradient decay. For the local cost, the variance lies around the $10^{-4}$ and varies only moderately, peaking near 384 trainable parameters and decreasing gently towards 640. Crucially, in all cases the local-cost variance stays well-separated from numerical noise and systematically higher than the global-cost variance, indicating that simply increasing the number of trainable parameters within this range does not trigger an immediate collapse of local gradients. 

Finally, Fig. \ref{figure6}c presents the layer-wise mean gradient variance for the exact architecture used in our experiments (16 qubits, 256 encoding gates, 512 trainable parameters, and 4 entangling layers). During circuit generation, we keep track of which trainable column (layer index) each parameter belongs to and then aggregate the variance per layer over the ensemble. For the global cost, all layers exhibit very small variances ($\sim 10^{-6}$–$10^{-5}$), as expected from Fig. \ref{figure6}a, but the profile is not monotonically collapsing with depth. For the local cost, the per-layer variance remains on the order of $10^{-4}$ over a broad band of intermediate layers and only decays modestly near the deepest part of the circuit. We do not observe a pathological pattern in which only a handful of final layers receive signal while earlier layers are effectively dead; instead, a substantial fraction of the depth contributes non-trivial gradients at random initialization.

Overall, the statevector diagnostics show that if we paired DAQC ansatz with a fully global observable such as $Z^{\otimes n}$, we would indeed encounter an exponentially vanishing gradient as number of qubits grows, i.e., a standard global barren plateau. For the local-type costs relevant to our classifier, gradient variances scale much more benignly with qubit number and remain significantly larger than in the global case over the range $n \le 22$. Moreover, at the chosen operating point (16 qubits, 256 encoding gates, 512 trainable parameters), gradients at random initialization are small but clearly non-vanishing for the local cost, and they are distributed across many layers rather than being confined to a vanishingly thin slice of the circuit.

\subsubsection{Gradient dynamics during training}

\begin{figure}[!ht]
  \centering
  \begin{subfigure}[t]{0.5\linewidth}
    \centering
    \includegraphics[width=\linewidth]{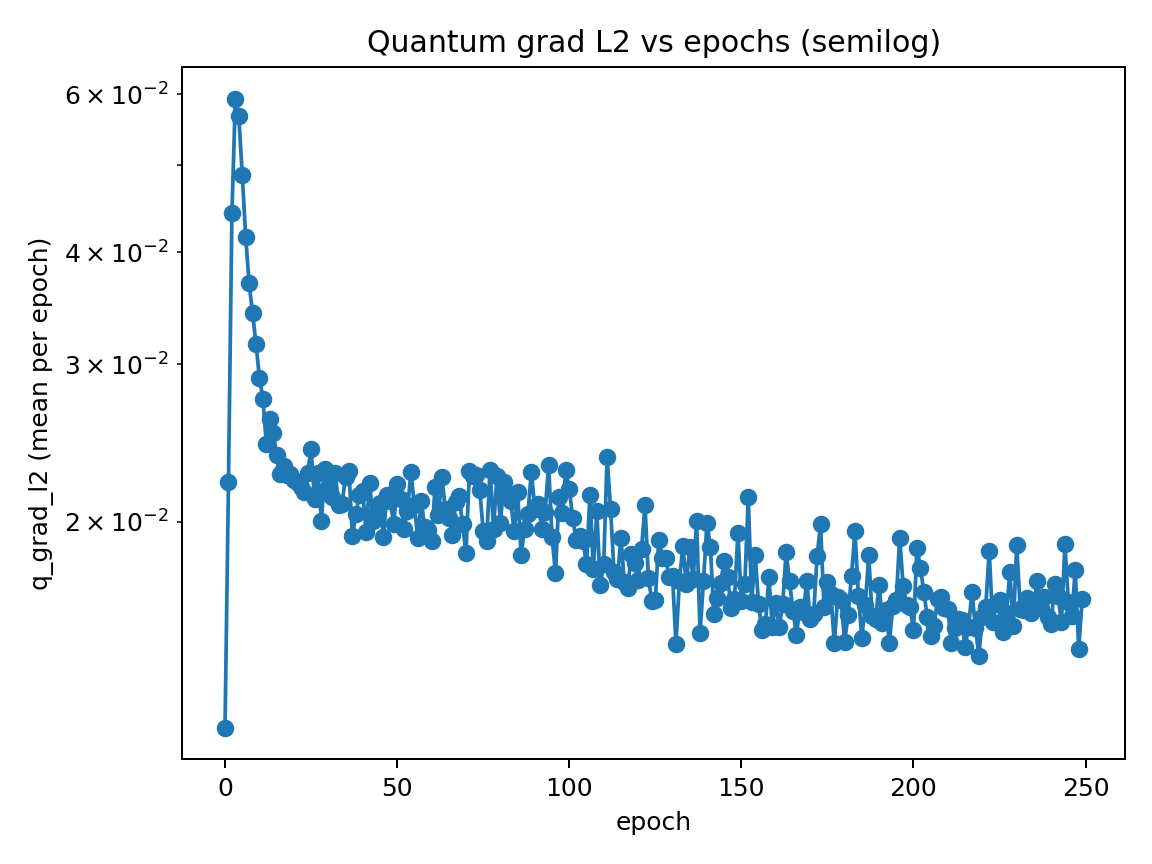}
    \caption{}
    \label{fig:daqc-a}
  \end{subfigure}\hfill
  \begin{subfigure}[t]{0.5\linewidth}
    \centering
    \includegraphics[width=\linewidth]{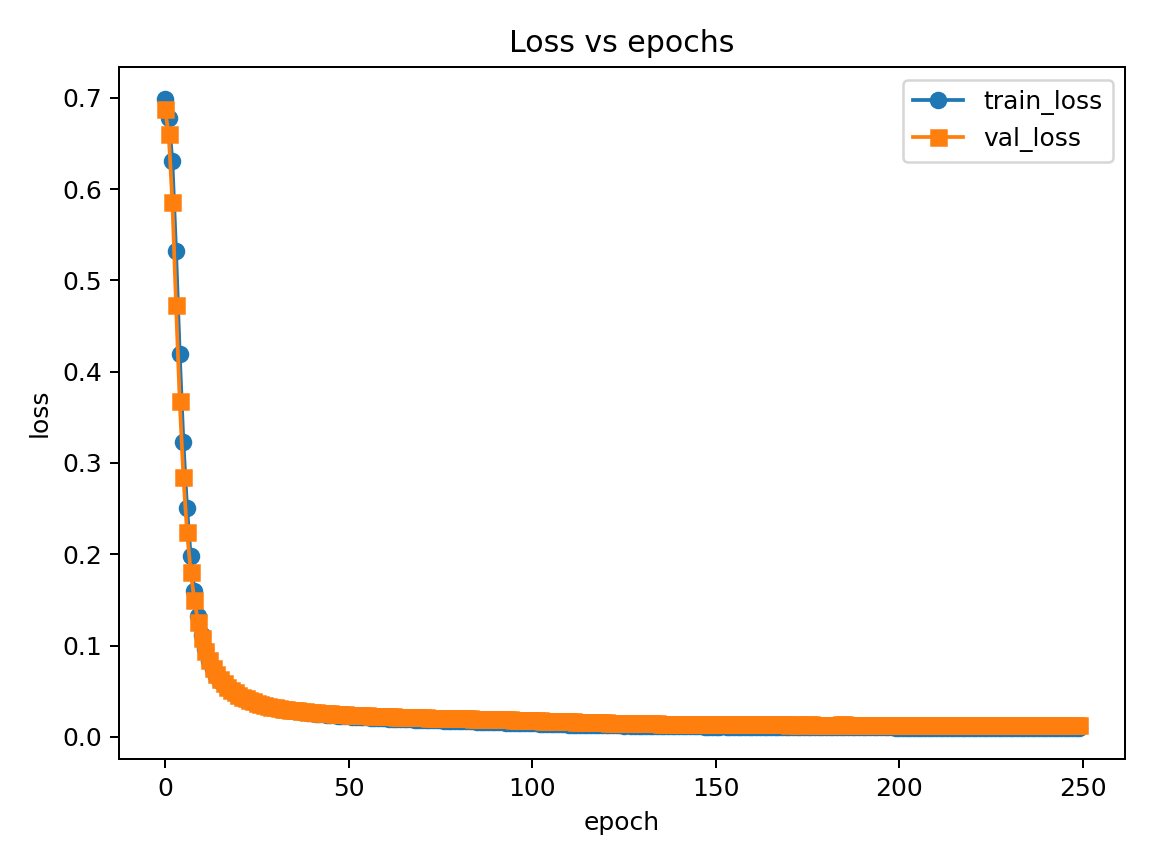}
    \caption{}
    \label{fig:daqc-b}
  \end{subfigure}\hfill
\caption{Gradient statistics and training dynamics for the 16-qubit DAQC on MNIST-2.
(a) Mean L2 norm of the gradients over quantum parameters versus epochs and (b) Training and validation loss versus epochs.}
\label{figure7}
\end{figure}

To verify that gradients remain usable along the training trajectory, we track the gradient statistics of the trainable quantum parameters during supervised learning on MNIST-2 dataset for the 16-qubit DAQC used in our experiments. Concretely, after each backward pass (and before the parameter update) we collect all gradients associated with the trainable quantum gates and compute their L2 norm. These per-step norms are accumulated in per-epoch buckets and averaged to obtain an epoch-wise mean quantum gradient norm. In parallel, we record the average training and validation cross-entropy losses at the end of each epoch. Fig. \ref{figure7} summarize these statistics over 250 epochs of training.

Fig. \ref{figure7}a reports the epoch-wise mean quantum gradient L2 norm on a semi-logarithmic $y$-axis. In the first few epochs the norm decreases to $2\times 10^{-2}$ range as the model moves away from random initialization. Crucially, it does not continue to drift downward toward numerical noise. Instead, after this initial transient the gradient norm stabilizes in a relatively narrow band around $\sim 2\times 10^{-2}$ for the remaining 200+ epochs, with only mild fluctuations and there is no signature of exponential decay. Fig. \ref{figure7}b shows the training and validation losses versus epochs. Both
curves drop rapidly from $\sim 0.7$ at initialization to below $0.05$ within the first few tens of epochs and then flatten into a smooth, low-loss plateau. The close tracking between train and validation curves indicates stable optimization without overfitting.

This behavior is exactly what we want to see if the circuit is practically trainable: gradients shrink as the model learns, but do not collapse. Combined with the local-cost scaling from the McClean-style analysis, the training-time statistics support a consistent picture: for the 16-qubit DAQC (256 embedding gates, 512 trainable parameters, and 64 two-qubit gates) with local cross-entropy objective, we do not encounter a practically limiting  barren plateau. Beyond this scale, barren plateaus may well reappear, but within the problem size considered, the combination of local measurements and hardware-aware design keeps the optimization landscape numerically accessible.

\section{Conclusion}

This work introduced a domain- and device-aware quantum circuit (DAQC) co-designed with (i) image-pixel locality and (ii) the heavy-hex connectivity of IBM backends. DAQC is expressive enough to learn non-trivial multi-class decision boundaries, while constraining two-qubit depth and limiting SWAP overhead. The unique design choices of DAQC also helps to tackle the barren plateau issue. The representational power of DAQC is consistently validated across MNIST, FashionMNIST, and PneumoniaMNIST using only $16\times16$ inputs and a few hundred trainable parameters. On the easier 2- and 4-class tasks, DAQC achieves performance that is competitive with strong classical baselines that operate on full $28\times28$ inputs with millions of parameters, and this behavior largely persists when executing on real IBM hardware with error mitigation. On the harder 10-class tasks, DAQC maintains high AUC (around 0.95), indicating that it can extract useful decision signals with few parameters and input constraints. On the imbalanced PneumoniaMNIST-2 dataset, DAQC attains a favorable trade-off between sensitivity and specificity compared to the classical models. Against recent QCS baselines, DAQC delivers substantially higher accuracy, F1-score, and more balanced sensitivity–specificity, highlighting the value of domain-aware, hardware-aligned circuit design over generic circuit search. In future, we will explore the tensor networks based simulation to scale the design of DAQC and its applications to more complex datasets and QML tasks.

\subsection{Scalability and outlook toward quantum advantage}

Quantum advantage can be understood as a quantum computation that demonstrably offers better efficiency, cost-effectiveness, or accuracy than what is achievable with classical computation alone \cite{lanes2025framework}. In the context of this work, we stride toward advantage framed in terms of resource efficiency and representational capability, rather than raw accuracy alone. DAQC operates under strict constraints---\(16\times16\) pooled inputs, 16 logical qubits, a few hundred trainable parameters, and moderate entangling depth---yet achieves performance that is competitive with classical baselines that process full \(28\times28\) images with millions of parameters. Concretely, this comparison can be organized along two axes: (i) \emph{efficiency}, where DAQC closely matches classical baselines on the 2- and 4-class tasks and maintains high AUC on the 10-class task despite reduced input resolution and compact parameterization; and (ii) \emph{cost-effectiveness}, where a structured, locality-preserving layout and ring-ECR entanglement keep two-qubit gate counts and depths within realistic NISQ budgets. Here, cost-effectiveness is understood in terms of quantum resources: DAQC achieves these results with only 16 logical qubits, $\approx$ 160 two-qubit gates, and a two-qubit depth of roughly 150 after transpilation. Under this lens, our results provide practical efficiency gains that strong claims of quantum advantage may also demonstrate. For a fixed or even reduced resource budget, a carefully designed quantum model can attain competitive classification performance and can be deployed on current hardware with standard suppression and mitigation techniques.

DAQC also exhibits a controlled and interpretable scaling behavior. Our ablation study over the number of entangling layers shows that a modest number of ECR layers is sufficient to capture useful correlations, whereas pushing depth too far degrades performance, reflecting the broader difficulty of scaling QML circuits on noisy devices. Our expressibility and entangling-capability analysis indicates that a configuration with 16 qubits, 256 embedded features (16 embedding layers), 512 trainable parameters (32 trainable layers), and 64 ECR gates (4 entangling layers) strikes a favorable balance between expressivity and robustness. Embedding substantially more than 256 features within this architecture would typically require increasing the number of qubits and additional embedding layers, both of which enlarge the physical footprint and amplify noise and compilation overhead on current hardware. This suggests a natural scaling strategy: increase embedded features and variational depth cautiously, so that gains in expressivity and sensitivity are not outweighed by noise and transpilation-induced depth. As hardware improves in terms of two-qubit fidelities, connectivity, and available depth, the same domain- and device-aware design principles can be applied at larger scales, closing the remaining performance gap to strong classical baselines while preserving a favorable efficiency profile. DAQC can be seen as an early step toward QML models that are adopted not because they are quantum, but because,
for certain data regimes and hardware generations, they could offer an attractive trade-off between accuracy, parameter count, and quantum resource usage compared to the classical state of the art.

\section*{Acknowledgments}
The authors gratefully acknowledge financial support from the NIH (GM130641).  

\section*{Author contributions}
G. Singh proposed and designed the study of domain-aware quantum circuit for QML. G. Singh performed the experiments under the supervision of T. Pellegrini and K.M. Merz. G. Singh wrote the manuscript, with inputs and contributions from all authors.

\section*{Competing interests}
The authors declare no competing interests.

\bibliographystyle{unsrt}  
\bibliography{references}

\end{document}